\documentclass[11pt]{article}

\usepackage{graphics}

\begin{document}

\title{Sewing string tree vertices with ghosts}

\author{Leonidas Sandoval Junior\thanks{E-mail address: dma2lsj@dcc.fej.udesc.br}\\ Department
of Mathematics\\ Centro de Ci\^encias Tecnol\'ogicas\\ UDESC - Universidade do Estado de Santa
Catarina - Brazil}

\maketitle

\begin{abstract}
It is shown how to sew string vertices with ghosts at tree level in order to produce new tree vertices using the Group Theoretic approach to String Theory. It is then verified the BRST invariance of the sewn vertex and shown that it has the correct ghost number.
\end{abstract}

\begin{flushright}\noindent PACS number: 11.25\end{flushright}

\vskip 0.5 cm

\section{Introduction}

In the early days of String Theory, one way to obtain amplitudes for the scattering of an arbitrary number of strings was by using the factorization property, what means that the scattering amplitude of $N$ strings may be interpreted as the scattering amplitudes of a smaller number of strings sewn together. This made it possible to build the $N$ string scattering amplitude by knowing the expression for three string scattering amplitudes. Even though they were very ingenious and successful, those calculations didn't take into account the ghost structure of the vertices, and that is what is done here.

In \cite{1}, it was shown how to sew tree vertices without ghosts using the Group Theoretic
approach to String Theory \cite{2} in order to obtain a new, composite vertex. Following the same procedure, we shall calculate the scattering amplitude of $N$ strings taking account the ghost structure.

We shall start with a short review of how to sew tree vertices without ghosts. What we must do is sew two legs of two vertices, one leg from each vertex. What we have in the beginning are two vertices $V_1$ and $V_2$ with $N_1$ and $N_2$ legs, respectively (figure 1).

\vskip 0.4 cm 

\includegraphics{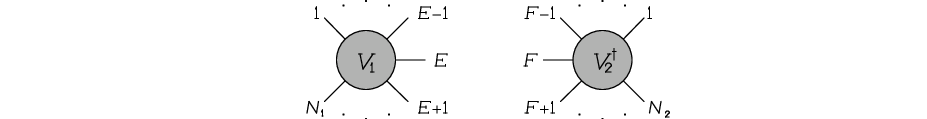}
\begin{center}
{\small Figure 1: Individual vertices.}
\end{center}

\vskip 0.2 cm 

We now sew leg $E$ from $V_1$ with the adjoint of leg $F$ from $V_2$. What we have now is the substitution of the two sewn legs by a propagator (figure 2). When this propagator is written in parametric form, it is an integration of one of the variables (in order to cancel one spurious degree of freedom) and a conformal factor ${\cal P}$ which contains terms of $L_n$'s acting on leg $E$ only.

\newpage 

\vskip 0.4 cm 

\includegraphics{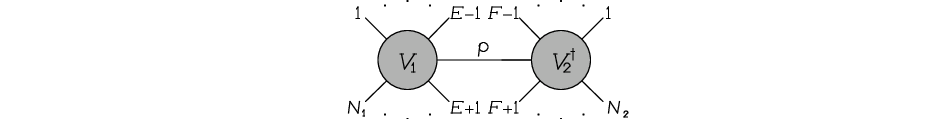}
\begin{center}
{\small Figure 2: Sewn vertices.}
\end{center}

\vskip 0.2 cm 

So the resulting vertex $V_c$ (called the {\sl composite vertex})  has the generic form
\begin{equation}
\label{eq4.1}
V_c=V_1PV_2^\dagger \ ,
\end{equation}
where the hermitian conjugate of $V_2$ is for the sewn leg $F$ only and
\begin{equation}
P=\int dx \ {\cal P}
\end{equation}
where $x$ is a suitable variable. In what follows, we shall often write ${\cal P}$ instead of $P$, calling attention to the integration when necessary.

When the two vertices are sewn together, we identify legs $E$ and $F$. We also identify the
Koba-Nielsen variable $z_E$ with one of the Koba-Nielsen variables of vertex $V_2^\dagger $,
and the Koba-Nielsen variable $z_F$ we identify with one of the Koba-Nielsen variables of
vertex $V_1$. In \cite{1}, this identification is made in the following way: $z_E$ may be
identified with $z_{F-1}$ or $z_{F+1}$, and $z_F$ may be identified with $z_{E-1}$ or
$z_{E+1}$. So there are four possible combinations: a) $z_E=z_{F-1}\ ,\ z_F=z_{E-1}$; b)
$z_E=z_{F+1}\ ,\ z_F=z_{E+1}$; c) $z_E=z_{F-1}\ ,\ z_F=z_{E+1}$; d) $z_E=z_{F+1}\ ,\
z_F=z_{E-1}$.

\section{Oscillator case}

The two original vertices satisfy some overlap identities and so shall do the composite vertex. One particular overlap identity is given by considering the operator $Q^{\mu i}$ with conformal weight $d=0$ defined by \cite{2}
\begin{equation}
Q^{\mu i}(\xi _i)=-\sum_{\scriptstyle n=-\infty \atop \scriptstyle n\neq 0}^\infty
\frac{1}{\sqrt{n}}\alpha ^{\mu i}_n(\xi _i)^{-n}+\alpha ^{\mu i}_0\ln \xi _i
+\frac{\partial \!\!\!\!{ }^{{ }^\leftarrow }}{\partial \alpha ^i_{0\mu }}\ \ .
\end{equation}
where $\alpha ^{\mu i}_n$ are bosonic oscillators with commutation relations
\begin{eqnarray}
\label{eq4.2}
 & & [\alpha ^{\mu i}_n,a^{\nu j}_m]=-\eta ^{\mu \nu }\delta ^{ij}\delta _{n,-m}\ ,
\ n,m\neq 0\ ,\\
\label{eq4.3}
 & & [\alpha ^{\mu i}_0,a^{\nu j}_n]=0\ ,\ \forall n\ .
\end{eqnarray}
The overlap identity is given by
\begin{equation}
V\left[ Q^{\mu i}(\xi _i)-Q^{\mu j}(\xi _j)\right] \ .
\end{equation}

Because we are considering the adjoint of leg $F$ in vertex $V_2$, we must see what the
adjoint of these overlap identities are. First, by the definition of $Q^{\mu i}$, we have that
\begin{equation}
\label{eq4.1a}
Q^{\mu i\dagger }(\xi _i)=\Gamma Q^{\mu i}(\xi _i)\Gamma =Q^{\mu i}(\xi _i)\ .
\end{equation}
So, the adjoint of the overlap equations is given by
\begin{equation}\left[ Q^{\mu i}(\Gamma \xi _i)-Q^{\mu j}(\xi _j)\right] V^\dagger =0\ .
\end{equation}

We now take the overlap identity considering the effects of the operator $Q^{\mu i}(\xi _i)$
on the vertex $V_1$ on a generic leg $i$ and on leg $E$\footnote{As we shall be seeing soon,
this form of the overlap will not lead to the correct composite vertex in the case where the
cycling transformations of the legs that are not sewn involve the sewn legs $E$ or $F$.}
(figure 3):

\hskip 0.7 cm \includegraphics{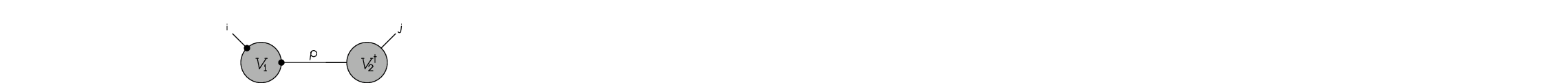}
\begin{center}
{\small Figure 3: Overlap identity for $V_1$.}
\end{center}

\begin{equation}
V_1\left[ Q^{\mu i}(\xi _i)-Q^{\mu E}(\xi _E)\right] =0\ .
\end{equation}

\vskip 0.2 cm
We may then insert the unit operator $1={\cal P}{\cal P}^{-1}$ and multiply by ${\cal P}$
without altering the result:
\begin{equation}
V_1{\cal P}{\cal P}^{-1}\left[ Q^{\mu i}(\xi _i)-Q^{\mu E}(\xi E)\right] {\cal P}\ .
\end{equation}

Since the conformal operator ${\cal P}$ acts only on leg $E$, it will have no effect on
$Q^{\mu i}(\xi _i)$. In order to compute the effect of ${\cal P}$ on $Q^{\mu E}(\xi _E)$, we
must know that, for a conformal transformation $V$ acting on a conformal operator $R(z)$ of
weight $d$,
\begin{equation}
VR(z)V^{-1}=\left( \frac{dVz}{dz}\right) ^dR(z)\ .
\end{equation}
Since $Q^{\mu E}(\xi _E)$ has conformal weight $d=0$, we have
\begin{equation}
{\cal P}^{-1}Q^{\mu E}(\xi _E){\cal P}=Q^{\mu E}\left( {\cal P}^{-1}\xi _E\right) \ ,
\end{equation}
and (figure 4)

\hskip 0.7 cm \includegraphics{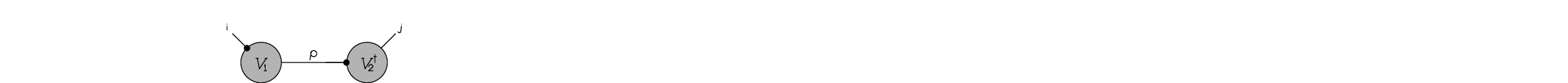}
\begin{center}
{\small Figure 4: Overlap identity for $V_1{\cal P}$.}
\end{center}

\begin{equation}
V_1{\cal P}\left[ Q^{\mu i}{\xi _i}-Q^{\mu E}({\cal P}^{-1}\xi _E)\right] =0\ .
\end{equation}
\vskip 0.2 cm

The second term in the expression above is facing now leg $F$ of vertex $V_2^\dagger $, or
best, its Hermitian conjugate. Considering that the Hermitian conjugate of $Q^{\mu i}$ (given
by (\ref{eq4.1a})), we then have the following overlap identity between legs $i$ and $F$:
\begin{equation}
V_1{\cal P}\left[ Q^{\mu i}(\xi _i)-Q^{\mu F}(\Gamma {\cal P}^{-1}\xi _F)\right] =0\ .
\end{equation}
We can then make a cycling transformation in order to obtain the correct factor for an
arbitrary leg $j\ (j\neq F)$ of vertex $V_2^\dagger $. The only term that will be affected is
the term depending on leg $F$:
\begin{equation}
V_j^{-1}V_FQ^{\mu F}\left( \Gamma {\cal P}^{-1}\xi _E\right) V_F^{-1}V_j=Q^{\mu j}
\left( V_j^{-1}V_F\Gamma {\cal P}^{-1}\xi _E\right) \ .
\end{equation}
Doing this, the overlap identity for the composite vertex $V_c$ (figure 5) can be written as

\hskip 3.1 cm \includegraphics{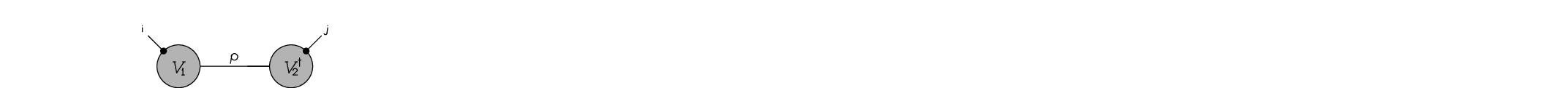}
\begin{center}
{\small Figure 5: Overlap identity for $V_c$.}
\end{center}

\begin{equation}
V_c\left[ Q^{\mu i}(\xi _i)-Q^{\mu j}(V^{-1}_jV_F\Gamma {\cal P}^{-1}\xi _E)\right] =0\ ,
\end{equation}
\vskip 0.2 cm
\noindent which is the overlap equation between two arbitrary legs $i$ and $j$ of the composite
vertex $V_c$.

But the overlap identity for the composite vertex, since none of legs $i$ or $j$ involves the
propagator, must be given by
\begin{equation}
V_c\left[ Q^{\mu i}(\xi _i)-Q^{\mu j}(\xi _j)\right] =0
\end{equation}
and so in order for the equation we have obtained for the overlap of the composite vertex
$V_c$ to be true we must have
\begin{equation}
V_F\Gamma {\cal P}^{-1}V^{-1}_E=1\Rightarrow {\cal P}^{-1}=\Gamma ^{-1}V^{-1}_FV_E
\end{equation}
which implies that the propagator is given by
\begin{equation}
\label{eq4.6}
{\cal P}=V^{-1}_EV_F\Gamma \ .
\end{equation}

In order to give an explicit expression for the propagator, we will now choose $\xi _i$ to be
of the form
\begin{equation}
\label{eq4.7}
\xi _i=V_i^{-1}z=z-z_i\ .
\end{equation}
This choice is called the ``simple cycling'' \cite{1} and it is the one that simplifies our
calculations the most. In this choice, the propagator is given by
\begin{equation}
{\cal P}z=\frac{1}{z}+z_F-z_E\ ,
\end{equation}
or in terms of the $L_n^E$ operators\footnote{Other forms for this propagator are given by
\cite{1}:
\begin{eqnarray*}
{\cal P} & = & {\rm e}^{-L_1^E/s}(-1)^{L_0^E}s^{2L_0^E}{\rm e}^{L_{-1}^E/s}\\
{\cal P} & = & (-1)^{L_0^E-L_1^E/s}s^{2(L_0^E-L_1^E/s)}{\rm e}^{-(L_1^E-L_{-1}^E)/s}\ .
\end{eqnarray*}},
\begin{equation}
\label{eq4.8}
{\cal P}={\rm e}^{(z_F-z_E-1)L_{-1}^E}(-1)^{L_0^E}{\rm e}^{L_1^E}{\rm e}^{-L_{-1}^E}\ .
\end{equation}

This form works for all choices for the composite vertex discussed before\footnote{This
affirmation is usually not valid for other choices of $\xi _i$.}. The true propagator is given
by expression (\ref{eq4.8}) integrated over a suitable variable. Choosing this variable to be
$s=z_F-z_E$, we then have
\begin{equation}
\label{eq4.9}
P=\int_{-\infty }^0 ds\ {\cal P}=\frac{1}{L_{-1}^E}(-1)^{L_0^E}{\rm e}^{L_1^E}
{\rm e}^{-L_{-1}^E}\ \ ,\ \ s=z_F-z_E\ .
\end{equation}

Before going any further, we must discuss another aspect of the theory that depends on the
particular way in which the legs are identified during the sewing procedure. Let us consider
the more general case of an arbitrary cycling $V_i$. This kind of cycling may depend on other
 coordinates that are not $z_i$. As an example, let us suppose that we are identifying
coordinate $z_E$ of leg $E$ with coordinate $z_{F-1}$ of vertex $V_2^\dagger $ and coordinate
$z_F$ of leg $F$ with coordinate $z_{F-1}$ of vertex $V_1$. The overlap identity between legs
$i$ and $E-1$ on vertex $V_1$ is
\begin{equation}
V_1\left[ Q^{\mu i}\left( V_{0i}^{-1}z\right) -Q^{\mu (E-1)}\left( V_{0(E-1)}^{-1}z\right)
\right]
\end{equation}
where we are calling $V_{0j}^{-1}$ the cycling transformation on leg $j$ $(j=1,\dots ,N)$.
These cycling transformations may depend on the other legs. As an example when this happens,
we take another choice of the cycling transformations $\xi _i$ that is not as trivial as
(\ref{eq4.7}) but gives a simpler formula for the propagator. This choice is given by \cite{1}
\cite{3}
\begin{equation}
\label{eq4.10}
\xi _i=V_{0i}^{-1}z=\frac{(z_{i+1}-z_{i-1})}{(z_{i+1}-z_i)}\cdot \frac{(z-z_i)}{(z-z_{i-1})}
\end{equation}
which is the transformation that takes $z_{i-1}$, $z_i$ and $z_{i+1}$ to $\infty $, 0 and 1,
respectively. Its inverse is given by
\begin{equation}
V_{0i}z=\frac{z_{i-1}(z_i-z_{i+1})z+z_i(z_{i+1}-z_{i-1})}{(z_i-z_{i+1})z+(z_{i+1}-z_{i-1})}\ .
\end{equation}

In the case of the cycling given by (\ref{eq4.10}), the cycling for leg $E-1$ will depend on
leg $E$, which is not present in the composite vertex. On this vertex, the overlap between
leg $E-1$ and an arbitrary leg $j$ reads
\begin{equation}
V_c\left[ Q^{\mu (E-1)}\left( V_{0(E-1)}^{-1}z\right) -Q^{\mu j}\left( V^{-1}_{0j}z\right)
\right] \ .
\end{equation}
This overlap equation involves terms that depend on leg $E$, which is non-existent in the
composite vertex $V_c$. The correct overlaps should be given by
\begin{equation}
V_c\left[ Q^{\mu (E-1)}\left( V_{E-1}^{-1}z\right) -Q^{\mu j}\left( V^{-1}_jz\right) \right]
\ ,
\end{equation}
where the cycling transformations $V_{E-1}^{-1}$ and $V^{-1}_j$ do not depend on legs $E$ or
$F$. So, in order to restore the correct cycling transformation for the composite vertex, a
conformal transformation must be made on the cycling transformations on vertex $V_1$. These are
given by
\begin{equation}
V_{0j}^{-1}\longrightarrow C_1V_{0j}^{-1}\ \ ,\ \ j\neq E\ ,
\end{equation}
where
\begin{equation}
\label{eq4.11}
C_1=\prod_{i=1}^{N_1}V_i^{-1}V_{0i}\ .
\end{equation}
In this definition, we consider implicit that the transformation $V_E^{-1}V_{0E}=1$ since the
conformal transformations on leg $E$ will not be part of the composite vertex and so need not
be modified.

Considering the general case, we have that the overlap equation for $V_1$ (figure 6) that will
lead to the correct composite vertex is now obtained from the original overlap

\hskip 0.8 cm \includegraphics{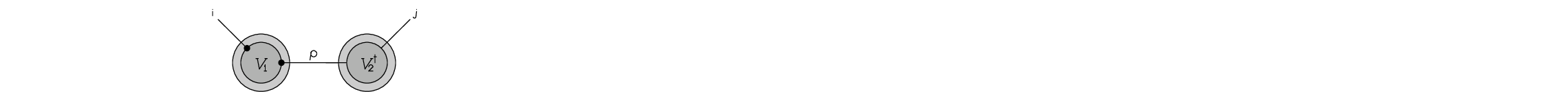}
\begin{center}
{\small Figure 6: Overlap identity for $V_1$.}
\end{center}

\begin{equation}
V_1\left[ Q^{\mu i}(\xi _{0i})-Q^{\mu E}(\xi _{0E})\right] =0\ ,
\end{equation}
where $\xi _{0i}=V_{0i}^{-1}z$ and $\xi _{0E}=V_{0E}^{-1}z$. By inserting conformal
transformation (\ref{eq4.11}), we obtain (figure 7)

\hskip 0.8 cm \includegraphics{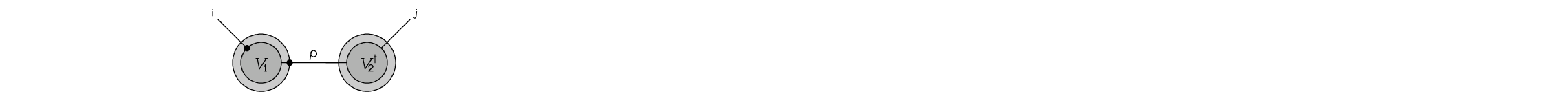}
\begin{center}
{\small Figure 7: Overlap identity for $V_1C_1^{-1}$.}
\end{center}

\begin{equation}
V_1C_1^{-1}\left[ Q^{\mu i}(\xi _i)-Q^{\mu E}(\xi _E)\right] =0\ .
\end{equation}
Inserting now the propagator, we obtain (figure 8)

\newpage 

\hskip 0.8 cm \includegraphics{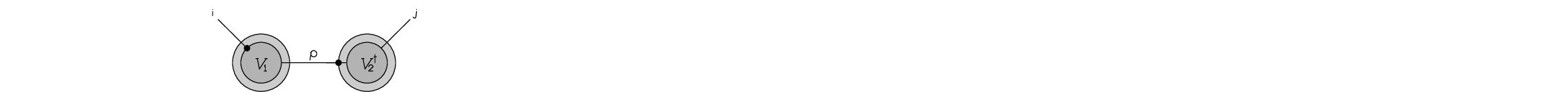}
\begin{center}
{\small Figure 8: Overlap identity for $V_1C_1^{-1}{\cal P}$.}
\end{center}

\begin{equation}
V_1C_1^{-1}{\cal P}\left[ Q^{\mu i}(\xi _i)-Q^{\mu E}({\cal P}^{-1}\xi _E)\right] =0
\end{equation}
and we expect the composite vertex to have a different form (given shortly) than in
(\ref{eq4.1}) in order to amount for the contributions of the conformal transformations. The
second term of the overlap is now facing leg $F$ of vertex $V_2^\dagger $ so that we have the
following overlap between legs $i$ and $F$:
\begin{equation}
V_1C_1^{-1}{\cal P}\left[ Q^{\mu i}(\xi _i)-Q^{\mu F}(\Gamma {\cal P}^{-1}\xi _E)\right] =0\ .
\end{equation}
We are now facing the conformal transformation $C_F$\footnote{In this case, like we have seen
for leg $V_E$, $V_F$ may depend on the variable $z_E$.}, defined by
\begin{equation}
\label{eq4.12}
C_F=V_F^{-1}V_{0F}
\end{equation}
which is necessary in order to change $\xi _{0F}\rightarrow \xi _F$. Introducing this
transformation we obtain (figure 9)

\hskip 0.8 cm \includegraphics{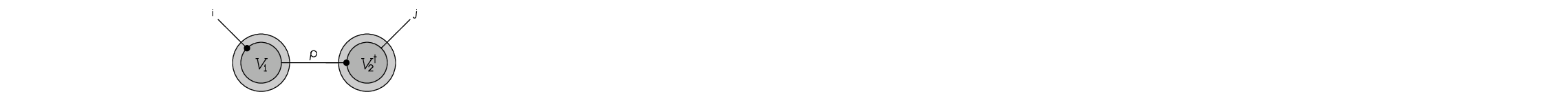}
\begin{center}
{\small Figure 9: Overlap identity for $V_1C_1^{-1}{\cal P}C_F$.}
\end{center}

\begin{equation}
V_1C_1^{-1}{\cal P}C_F\left[ Q^{\mu i}(\xi _i)-Q^{\mu F}(V_{0F}^{-1}V_F\Gamma
{\cal P}^{-1}\xi _E)\right] =0\ .
\end{equation}

Making a cycling transformation from leg $F$ to leg $j$, we then obtain (figure 10)

\hskip 0.8 cm \includegraphics{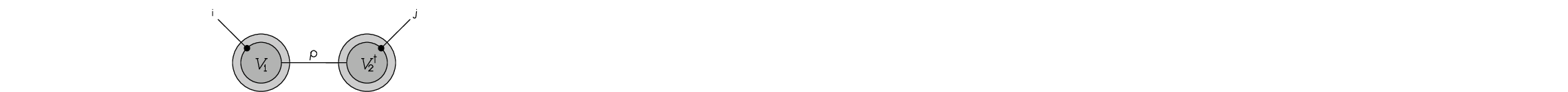}
\begin{center}
{\small Figure 10: Overlap identity for $V_1C_1^{-1}{\cal P}C_FV_2^\dagger $.}
\end{center}

\begin{equation}
V_1C_1^{-1}{\cal P}C_FV_2^\dagger \left[ Q^{\mu i}(\xi _i)-Q^{\mu j}(V_{0j}^{-1}V_F\Gamma
{\cal P}^{-1}\xi _E)\right] =0\ .
\end{equation}
Once again, a conformal transformation must be introduced because of the cycling transformations $V_{0j}^{-1}$. This is defined by
\begin{equation}
\label{eq4.13}
C_2=\prod_{\scriptstyle i=1\atop \scriptstyle i\neq F}^{N_2}V_i^{-1}V_{0i}\ .
\end{equation}
so that we now have (figure 11)

\newpage 

\hskip 0.8 cm \includegraphics{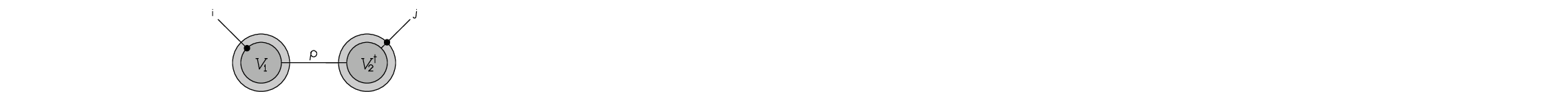}
\begin{center}
{\small Figure 11: Overlap identity for $V_1C_1^{-1}{\cal P}C_FV_2^\dagger C_2^{-1}$.}
\end{center}

\begin{equation}
V_1C_1^{-1}{\cal P}C_FV^\dagger _2C_2^{-1}\left[ Q^{\mu i}(\xi _i)-Q^{\mu j}(V^{-1}_jV_F
\Gamma {\cal P}^{-1}\xi _E)\right] =0\ .
\end{equation}

The composite vertex must be defined in terms of the new cycling transformations and so it
must now include the conformal transformations that perform this change. So, it will now be
defined by
\begin{equation}
\label{eq4.14}
V_c=V_1C_1^{-1}{\cal P}C_FV_2^\dagger C_2^{-1}\ .
\end{equation}
Considering this, the overlap identity for the composite vertex $V_c$ can be written as
\begin{equation}
V_c\left[ Q^{\mu i}(\xi _i)-Q^{\mu j}(V^{-1}_jV_F\Gamma {\cal P}^{-1}\xi _E)\right] =0\ .
\end{equation}
Since the correct overlap identity for the composite vertex is given by
\begin{equation}
V_c\left[ Q^{\mu i}(\xi _i)-Q^{\mu j}(\xi _j)\right] =0
\end{equation}
we must have
\begin{equation}
V_F\Gamma {\cal P}^{-1}V^{-1}_E=1
\end{equation}
which implies once again that
\begin{equation}
{\cal P}=V^{-1}_EV_F\Gamma \ .
\end{equation}

For the cycling transformation (\ref{eq4.10}), it is only necessary to do conformal
transformations on legs $E-1$, $E$, $E+1$, $F-1$, $F$ and $F+1$, depending on the particular
way the variables associated with these legs are identified with the variables associated to
legs $E$ and $F$. In this particular example (which is case $a$ seen before), the conformal
transformations are given by
\begin{eqnarray}
V_{0(E-1)}^{-1}\longrightarrow C_1V_{0(E-1)}^{-1}\ \  & ,\ \ & C_1=r^{L_0^{E-1}}\ ,\\
V_{0(F-1)}^{-1}\longrightarrow C_2V_{0(F-1)}^{-1}\ \  & ,\ \ & C_2=t^{L_0^{F-1}}
\end{eqnarray}
where
\begin{equation}
r=\frac{(z_{F-1}-z_{E-2})}{(z_{F-1}-z_{E-1})}\frac{(z_{E-1}-z_E)}{(z_{E-2}-z_E)}\ \ ,\ \
t=\frac{(z_{E-1}-z_{F-2})}{(z_{E-1}-z_{F-1})}\frac{(z_{F-1}-z_F)}{(z_{F-2}-z_F)}\ .
\end{equation}
For the cycling (\ref{eq4.10}), the propagators obtained for the four possible combinations
discussed before are given by
\begin{equation}
\label{eq4.15}
{\rm a)}\ \ {\cal P}_a=a^{L_0^E}\ \ ,\ \ {\rm b)}\ \ {\cal P}_b={\rm e}^{-L_1^E}a^{L_0^E}
{\rm e}^{-L_{-1}^E}\ \ ,\ \ {\rm c)}\ \ {\cal P}_c={\rm e}^{-L_1^E}b^{L_0^E}\ \ ,\ \
{\rm d)}\ \ {\cal P}_d=b^{L_0^E}{\rm e}^{-L_{-1}^E}\ ,
\end{equation}
where\footnote{The coefficient $a$ can be connected with the coefficient $c$ in reference
\cite{1} by $a=\frac{c}{c-1}$.}
\begin{equation}
a=\frac{(z_{E+1}-z_{E-1})(z_{F+1}-z_{F-1})}{(z_{E+1}-z_{F-1})(z_{F+1}-z_{E-1})}\ \ ,\ \
b=\frac{(z_{E+1}-z_{E-1})(z_{F+1}-z_{F-1})}{(z_{E+1}-z_{F+1})(z_{F-1}-z_{E-1})}\ .
\end{equation}
The true propagators are obtained when we integrate the expressions above multiplied by a
suitable constant. The results are:
\begin{eqnarray}
 & & {\rm a)}\ \ P_a=\int_0^1 da\ a^{L_0^E-1}=\frac{1}{L_0^E}\ \ ,\ \ {\rm b)}\ \ P_b=
\int_0^1da\ {\rm e}^{-L_1^E}a^{L_0^E-1}{\rm e}^{-L_{-1}^E}={\rm e}^{-L_1^E}\frac{1}{L_0^E}
{\rm e}^{-L_{-1}^E},\\
 & & {\rm c)}\ \ P_c=\int_0^1db\ {\rm e}^{-L_1^E}b^{L_0^E-1}={\rm e}^{-L_1^E}\frac{1}{L_0^E}
\ \ ,\ \ {\rm d)}\ \ P_d=\int_0^1db\ b^{L_0^E-1}{\rm e}^{-L_{-1}^E}=\frac{1}{L_0^E}
{\rm e}^{-L_{-1}^E}.
\end{eqnarray}

It is now necessary to verify the effect of the gauge transformations $C_1$ and $C_2$ on the
composite vertex as given by formula (\ref{eq4.14}). We shall do it by verifying the effect of
$C_1$ on vertex $V_1$. In order to do this we need the explicit expression for the bosonic
oscillator vertex $V_1$, given by \cite{1} \cite{4}
\begin{equation}
\label{eq4.19}
V_1=\left( \prod_{i=1}^N{ }_i\langle 0|\right) \exp \left[ -\frac{1}{2}\sum_{\scriptstyle
i,j=1\atop \scriptstyle i\neq j}^{N_1}\sum_{n,m=0}^\infty \alpha ^{\mu i}_nD_{nm}
\left( \Gamma V^{-1}_{0i}V_{0j}\right) \alpha ^j_{m\mu }\right]
\end{equation}
where $V_{0i}^{-1}$ and $V_{0j}$ are cycling transformations involving leg $E$ and the
oscillators $\alpha _n^{\mu i}$ have commutation relations given by (\ref{eq4.2},
\ref{eq4.3}). Matrices $D_{nm}(\gamma )$ are defined in the following way \cite{3}:
\begin{eqnarray}
D_{n0}(\gamma ) & = & \frac{1}{\sqrt{n}}\left[ \gamma (0)\right] ^n\ ,\\
D_{nm}(\gamma ) & = & \left. \sqrt{\frac{m}{n}}\frac{1}{m!}\frac{\partial ^m}
{\partial z^m}\left[ \gamma (z)\right] ^n\right| _{z=0}\ ,\\
D_{00}(\gamma) & = & \left. \frac{1}{2}\ln \left[ \frac{d}{dz}\gamma (z)\right] \right| _{z=0}
\end{eqnarray}
and have the following multiplication property:
\begin{equation}
\label{eq4.20}
D_{nm}(\gamma _1\gamma _2)=\sum_{p=1}^\infty D_{np}(\gamma _1)D_{pm}(\gamma _2)+D_{n0}
(\gamma _1)\delta _{0m}+\delta _{0n}D_{0m}(\gamma _2)\ .
\end{equation}

In order to calculate the effects of the conformal transformation $C_1$ on these oscillators
we must make use of the following conformal operator \cite{2}
\begin{equation}
P^{\mu i}(\xi _{0i})=\sum_{\scriptstyle n=-\infty \atop n\neq 0}^\infty \sqrt{|n|}
\alpha _n^{\mu i}(\xi _{0i})^{-n-1}+\alpha _0^{\mu i}(\xi _{0i})^{-1}
\end{equation}
which has conformal weight one, what means it transforms like
\begin{equation}
P^{\mu i}(\xi _{0i})=\frac{d\xi _{0j}}{d\xi _{0i}}P^{\mu j}(\xi _{0j})\ .
\end{equation}

An oscillator $\alpha _n^{\mu i}\ (n\geq 1)$ can be expressed in terms of this conformal
operator in the following way:
\begin{equation}
\alpha _n^{\mu i}=\frac{1}{\sqrt{n}}\oint _{\xi _{0i}=0}d\xi _{0i}\ (\xi _{0i})^nP^{\mu i}
(\xi _{0i})\ .
\end{equation}
Acting on it with the conformal transformation $C_1$, we have
\begin{equation}
C_1\alpha _n^{\mu i}C_1^{-1}=\frac{1}{\sqrt{n}}\oint _{\xi _{0i}=0}d\xi _{0i}\ (\xi _{0i})^n
\left( \frac{d}{d\xi _{0i}}V_i^{-1}V_{0i}\xi _{0i}\right) P^{\mu i}(V_i^{-1}V_{0i}\xi _{0i})\ .
\end{equation}
Making now a change of variables $\xi _{0i}\rightarrow \xi _i$, we have
\begin{equation}
C_1\alpha _n^{\mu i}C_1^{-1}=\frac{1}{\sqrt{n}}\oint _{\xi _i=0}d\xi _i\ (V_{0i}^{-1}
V_i\xi _i)^nP^{\mu i}(\xi _i)\ .
\end{equation}
Expanding $(V_{0i}^{-1}V_i\xi _i)^n$ in terms of $D_{nm}(\gamma )$ matrices and
$P^{\mu i}(\xi i)$ in terms of the oscillators, we obtain
\begin{eqnarray}
 & & C_1\alpha _n^{\mu i}C_1^{-1} =\nonumber \\
 & & \sum_{\scriptstyle p=-\infty \atop \scriptstyle p\neq 0}^\infty \sqrt{\frac{p}{n}}
\oint _{\xi _i=0}d\xi _i\ \left[ \sqrt{n}D_{n0}\left( V_{0i}^{-1}V_i\right) (\xi _i)^{-p-1}
+\sum_{m=1}^\infty \sqrt{\frac{n}{m}}D_{nm}\left( V_{0i}^{-1}V_i\right) (\xi _i)^{m-p-1}\right]
\alpha ^{\mu i}_p \nonumber \\
 & & +\frac{1}{\sqrt{n}}\oint _{\xi _i=0}d\xi _i\ \left[ \sqrt{n}D_{n0}\left( V_{0i}^{-1}V_i
\right) (\xi _i)^{-1} +\sum_{m=1}^\infty \sqrt{\frac{n}{m}}D_{nm}\left( V_{0i}^{-1}V_i\right)
(\xi _i)^{m-1}\right] \alpha ^{\mu i}_0\ .
\end{eqnarray}
Performing the integrations we then obtain
\begin{equation}
C_1\alpha _n^{\mu i}C_1^{-1}=\sum_{m=0}^\infty D_{nm}\left( V_{0i}^{-1}V_i\right)
\alpha _m^{\mu i}\ .
\end{equation}
Using the same process for $\alpha _0^{\mu i}$, we obtain
\begin{equation}
C_1\alpha _0^{\mu i}C_1^{-1}=\alpha _0^{\mu i}\ .
\end{equation}

Using these transformation properties, the multiplication rules of matrices $D_{nm}(\gamma )$
(equation (\ref{eq4.20})) and the property \cite{3}
\begin{equation}
D_{nm}(\gamma )=D_{mn}(\Gamma \gamma ^{-1}\Gamma )\ ,
\end{equation}
we can show that the effect of the conformal transformation $C_1$ on vertex $V_1$ is given by
\begin{equation}
\label{eq4.21}
V_1C_1^{-1}=\left( \prod_{i=1}^N{ }_i\langle 0|\right) \exp \left[ -\frac{1}{2}
\sum_{\scriptstyle i,j=1\atop \scriptstyle i\neq j}^{N_1}\sum_{n,m=0}^\infty \alpha ^{\mu i}_n
D_{nm}\left( \Gamma V^{-1}_iV_j\right) \alpha ^j_{m\mu }\right] \ ,
\end{equation}
i.e. the effect of $C_1$ on vertex $V_1$ is to change $V_{0i}^{-1}\rightarrow V_i^{-1}$ and
$V_{0j}\rightarrow V_j$ thus eliminating the dependence of the cycling transformations of every
leg except leg $E$ on the latter. The same can be done to obtain the effect of conformal
transformation $C_2$ on vertex $V_2^\dagger $, with the same results. So, the effect of these
transformations is to eliminate from the cycling transformations of the composite vertex all
dependence on the sewn legs $E$ and $F$.

\section{Introduction of ghosts}

We now introduce ghosts in the vertex so that what we must sew now are two vertices with some
ghost variables attached to them, i.e. we will be considering the physical vertices \cite{4}
which have the correct ghost number. In this case, in addition to satisfying the overlap
identities with the conformal operator $Q^{\mu i}$, the two physical vertices and the composite
vertex must also satisfy some overlap identities with the conformal operators $b^i$ and $c^i$,
given by \cite{4}
\begin{eqnarray}
\label{eq4.22}
b^i(\xi _i) & = & \sum_{n=-\infty }^\infty b^i_{-n}(\xi _i)^{n-2}\ ,\\
\label{eq4.23}
c^i(\xi _i) & = & \sum_{n=-\infty }^\infty c^i_{-n}(\xi _i)^{n+1}
\end{eqnarray}
where $c_n^i$ and $b_n^i$ are ghost anticommuting oscillators with anticommutation relations
\begin{equation}
\{ c_n^i,b_m^j\} =\delta _{n,-m}\ .
\end{equation}
These operators have, respectively, conformal weights $2$ and $-1$, what means that they
transform like
\begin{eqnarray}
b^i(\xi _i) & = & \left( \frac{d\xi _j}{d\xi _i}\right) ^2b^j(\xi _j)\ ,\\
c^i(\xi _i) & = & \left( \frac{d\xi _j}{d\xi _i}\right) ^{-1}c^j(\xi _j)\ .
\end{eqnarray}

The overlap identities for a vertex $V$ with these operators are given by
\begin{eqnarray}
\label{eq4.24}
 & & V\left[ b^i(\xi _i)-\left( \frac{d\xi _j}{d\xi _i}\right) ^2b^j(\xi _j)\right] =0\ ,\\
\label{eq4.25}
 & & V\left[ c^i(\xi _i)-\left( \frac{d\xi _j}{d\xi _i}\right) ^{-1}c^j(\xi _j)\right] =0\ .
\end{eqnarray}

We shall be working here with overlap identities for the physical vertex $U$ \cite{4}, which
has the correct ghost number, instead of the overlap identities for vertex $V$. The physical
vertex is given by \cite{4}
\begin{equation}
\label{eq4.26}
U=V\prod _{\scriptstyle i=1\atop \scriptstyle i\neq a,b,c}^N\sum_{j=1}^N\sum_{n=-1}^\infty
e_n^{ij}b_n^j
\end{equation}
where $a,b,c$ are any three legs of the vertex and the matrix $e_n^{ij}$ is given by
\begin{equation}
\sum_{n=-1}^\infty e_n^{ij}{\cal L}^j_n=V_j^{-1}\partial _{z_i}V_j\
\end{equation}
where the cycling transformations are now defined on the complete generators ${\cal L}_n^i$ of
the conformal algebra of the bosonic oscillators and of the ghost oscillators. These vectors
$e_n^{ij}$ have the following property:
\begin{equation}
\label{eq4.27}
\frac{\partial V}{\partial z_i}=V\sum_{j=1}^N\sum_{n=-1}^\infty e_n^{ij}{\cal L}_n^j\ .
\end{equation}

In order to derive the overlap identity for the physical vertex $U$, we must multiply the
overlap identity for $V$ by the same factor as in equation (\ref{eq4.26}),
\begin{eqnarray}
 & & V\left[ b^i(\xi _i)-\left( \frac{d\xi _j}{d\xi _i}\right) ^2b^j(\xi _j)\right]
\left( \prod _{\scriptstyle k=1\atop \scriptstyle k\neq a,b,c}^N\sum_{l=1}^N
\sum_{n=-1}^\infty e_n^{kl}b_n^l\right) =0\ ,\\
 & & V\left[ c^i(\xi _i)-\left( \frac{d\xi _j}{d\xi _i}\right) ^{-1}c^j(\xi _j)\right]
\left( \prod _{\scriptstyle k=1\atop \scriptstyle k\neq a,b,c}^N\sum_{l=1}^N
\sum_{n=-1}^\infty e_n^{kl}b_n^l\right) =0
\end{eqnarray}
and pass it through the overlap identities, obtaining
\begin{eqnarray}
\label{eq4.28}
 & & U\left[ b^i(\xi _i)-\left( \frac{d\xi _j}{d\xi _i}\right) ^2b^j(\xi _j)\right] =0\ ,\\
 & & U\left[ c^i(\xi _i)-\left( \frac{d\xi _j}{d\xi _i}\right) ^{-1}c^j(\xi _j)\right]
\nonumber \\
\label{eq4.29}
 & & +V\sum_{\scriptstyle p=1\atop p\neq \scriptstyle a,b,c}^N(-1)^p\prod _{\scriptstyle k=1
\atop {\scriptstyle k\neq a,b,c\atop \scriptstyle k\neq p}}^N\sum_{l=1}^N\sum_{q=-1}^\infty
e_q^{kl}b_q^l\sum_{n=-1}^\infty \left[ e_n^{pi}(\xi _i)^{n+1}-\left( \frac{d\xi _j}{d\xi _i}
\right) ^{-1}e_n^{pj}(\xi _j)^{n+1} \right] =0\ .
\end{eqnarray}

From (\ref{eq4.29}) we can see that there will be an anomalous term in the $c^i$ overlap of
the physical vertex $U$ unless both legs $i$ and $j$ are precisely those legs ($a$, $b$ or $c$)
that do not have any ghosts attached to them. These ghosts which are attached to all the other
legs are responsible for the anomalous terms.

\subsection{Analysis of the ghost number}

Before going any further, it is necessary to make some considerations on the ghost number of
the composite vertex. As we shall see shortly, in the case where we perform the sewing with
ghosts included, using the physical vertices, the resulting composite physical vertex will not
have the correct ghost number unless we insert some extra ghosts in vertex $U_1$ before the
sewing takes place. Considering this, we shall define the composite vertex to be given by
\begin{equation}
\label{eq4.30}
U_c=U_1GPU_2^\dagger
\end{equation}
where $G$ are some extra ghosts that will be introduced in order to make vertex $U_c$ have the
correct ghost number and $P$ is the propagator (in its integrated form).

We must now analyze the ghost number of the composite vertex and of its parts in order to
calculate the ghost number that the extra ghosts $G$ must have. In order to do this, we shall
use the {\rm ghost number operator} $N_{\rm gh}$. For a vertex with $N$ legs, the ghost number
operator is defined by
\begin{equation}
N^{\rm gh}=\sum_{i=1}^N\left( \sum_{n=-1}^\infty c_{-n}^ib_n^i-\sum_{n=2}^\infty b_{-n}^i
c_n^i\right) \ .
\end{equation}
The reason why the ghost number operator is a sum from $i=1$ to $i=N$ is because there are
$N$ vacua that will annihilate the operators corresponding to each one of them. When acting
on the physical vertex, this operator gives a ghost number $N$, what is the correct ghost
number for a tree vertex with $N$ legs.

In the case of the composite vertex, it has $N_1+N_2-2$ legs (because it does not have legs
$E$ and $F$, which have been sewn together) and so it must have ghost number $(N_1+N_2-2)$. For
this vertex, the ghost number operator $N_c^{\rm gh}$ can be divided into two parts:
\begin{equation}
N^{\rm gh}_c=N^{\rm gh}_1+N^{\rm gh}_2
\end{equation}
where
\begin{eqnarray}
N^{\rm gh}_1 & = & \sum_{\scriptstyle i=1\atop \scriptstyle i\neq E}^{N_1}\left(
\sum_{n=-1}^\infty c_{-n}^ib_n^i-\sum_{n=2}^\infty b_{-n}^ic_n^i\right) \ ,\\
N^{\rm gh}_2 & = & \sum_{\scriptstyle i=1\atop i\neq F}^{N_2}\left( \sum_{n=-1}^\infty
c_{-n}^ib_n^i-\sum_{n=2}^\infty b_{-n}^ic_n^i\right) \ .
\end{eqnarray}
This ghost number operator will have the following effect on the composite vertex:
\begin{equation}
\label{eq4.31}
U_cN_c^{\rm gh}=U_c\left( N_1^{\rm gh}+N_2^{\rm gh}\right) =(N_1+N_2-2)U_c\ .
\end{equation}
Given formula (\ref{eq4.30}) for the composite vertex, we then have
\begin{equation}
\label{eq4.32}
U_cN_c^{\rm gh}=U_1N_1^{\rm gh}GPU_2^\dagger +U_1\left[ G,N_1^{\rm gh}\right] PU_2^\dagger
+U_1GPU_2^\dagger N_2^{\rm gh}\ .
\end{equation}

In order to calculate this, we must pay some attention to terms one and three of the right
hand side of the expression above. We know that
\begin{eqnarray}
\label{eq4.33}
 & & U_1\left( N_1^{\rm gh}+N_E^{\rm gh}\right) =N_1U_1\ ,\\
\label{eq4.34}
 & & U_2\left( N_2^{\rm gh}+N_F^{\rm gh}\right) =N_2U_2
\end{eqnarray}
where
\begin{eqnarray}
N^{\rm gh}_E & = & \sum_{n=-1}^\infty c_{-n}^Eb_n^E-\sum_{n=2}^\infty b_{-n}^Ec_n^E\ ,\\
N^{\rm gh}_F & = & \sum_{n=-1}^\infty c_{-n}^Fb_n^F-\sum_{n=2}^\infty b_{-n}^Fc_n^F\ .
\end{eqnarray}
Taking the Hermitian conjugate of equation (\ref{eq4.34}) only on leg $F$, we obtain
\begin{equation}
\label{eq4.35}
U_2^\dagger N_2^{\rm gh}+{N_F^{\rm gh}}^\dagger U_2^\dagger =N_2U_2^\dagger \ .
\end{equation}

Since for ${N_F^{\rm gh}}^\dagger $ (and for any arbitrary ghost number operator)
\begin{equation}
\label{eq4.36}
{N_F^{\rm gh}}^\dagger =-N_F^{\rm gh}+3
\end{equation}
we then obtain, substituting (\ref{eq4.33}), (\ref{eq4.35}) and (\ref{eq4.36}) into equation
(\ref{eq4.32}),
\begin{equation}
U_cN_c^{\rm gh} = (N_1+N_2-3)U_1GPU_2^\dagger +U_1\left[ G,N_1^{\rm gh}\right] PU_2^\dagger
-U_1N_E^{\rm gh}GPU_2^\dagger +U_1GPN_F^{\rm gh}U_2^\dagger \ .
\end{equation}
Passing $N_E^{\rm gh}$ through the extra ghosts $G$, we then obtain
\begin{eqnarray}
U_cN_c^{\rm gh}  & = &  (N_1+N_2-3)U_1GPU_2^\dagger +U_1\left[ G,N_1^{\rm gh}\right] PU_2^
\dagger \nonumber \\
\label{eq4.37}
 & & -U_1\left[ N_E^{\rm gh},G\right] PU_2^\dagger -U_1GPN_E^{\rm gh}U_2^\dagger
+U_1GPN_F^{\rm gh}U_2^\dagger \ .
\end{eqnarray}

We must now remember that, in the composite vertex, we identify every operator on leg
$E$ with operators on leg $F$ so that $N_E^{\rm gh}=N_F^{\rm gh}$. Doing this, the last
two terms in (\ref{eq4.37}) cancel and we obtain the following result:
\begin{equation}
U_cN_c^{\rm gh}=(N_1+N_2-3)U_1GPU_2^\dagger +U_1\left[ G,N_1^{\rm gh}+N_E^{\rm gh}\right]
PU_2^\dagger \ .
\end{equation}
The fact that $U_c$ has ghost number $N_1+N_2-2$ then implies that
\begin{equation}
\left[ G,N_1^{\rm gh}+N_E^{\rm gh}\right] =G\ ,
\end{equation}
i.e. the extra ghosts that must be introduced in vertex $U_1$ must have ghost number 1.
\footnote{This contrasts with references \cite{5} and \cite{6} where it is claimed that the
extra ghosts should have ghost number 1, 2 or 3, depending on the way one chooses legs $E$
and $F$ to have or not to have ghosts attached to them.}

There is an infinite number of combinations of ghosts that have ghost number 1. We could have
any linear combination of ghosts of the type $b$, $bcb$, $bcbcb$, etc. but it will prove to be
simpler to choose $G$ to be a combination of $b's$ only so that we may represent it as
\begin{equation}
\label{eq4.38}
G=\sum_{i=1}^{N_1}\sum_{n=-\infty }^\infty \alpha _n^ib_n^i
\end{equation}
where $\alpha _n^i$ are the coefficients of the linear combination. In order to determine the
correct linear combination, we must use some other conditions, like BRST invariance of the
scattering amplitude. This we shall see next.

\subsection{BRST invariance}

We must now impose that the scattering amplitude obtained from the composite vertex is BRST
invariant and check whether this condition is strong enough to determine $G$. The scattering
amplitude \cite{1} is obtained by acting with the composite vertex
\begin{equation}
U_c=U_1GPU_2^\dagger
\end{equation}
on a certain number of physical states ($|\chi _1\rangle |\chi _2\rangle \dots
|\chi _N\rangle $) and then by integrating over all variables $z_i$ ($i=1,\dots ,N_1+N_2$;
$i\neq E,F$):
\begin{equation}
W=\int \prod_{\scriptstyle i=1\atop \scriptstyle i\neq E,F}^{N_1+N_2}dz_i\ U_1GPU_2^\dagger
|\chi _1\rangle |\chi _2\rangle \dots |\chi _N\rangle \ .
\end{equation}
$P$ is the propagator in its integrated form and $G$ are the extra ghosts to be inserted in
$U_1$.

The action of the BRST charge $Q$ on this scattering amplitude is given by
\begin{eqnarray}
 & & WQ  = \int \prod_{\scriptstyle i=1\atop \scriptstyle i\neq E,F}^{N_1+N_2}dz_i\
\left[ U_1,Q\right] GPU_2^\dagger |\chi _1\rangle |\chi _2\rangle \dots |\chi _N\rangle
\nonumber \\
 & & +\int \prod_{\scriptstyle i=1\atop \scriptstyle i\neq E,F}^{N_1+N_2}dz_i\ U_1
\left[ G,Q\right] PU_2^\dagger |\chi _1\rangle |\chi _2\rangle \dots |\chi _N\rangle
+\int \prod_{\scriptstyle i=1\atop \scriptstyle i\neq E,F}^{N_1+N_2}dz_i\ U_1G\left[ P,Q\right]
U_2^\dagger |\chi _1\rangle |\chi _2\rangle \dots |\chi _N\rangle \nonumber \\
 & & +\int \prod_{\scriptstyle i=1\atop \scriptstyle i\neq E,F}^{N_1+N_2}dz_i\ U_1GP
\left[ U_2^\dagger ,Q\right] |\chi _1\rangle |\chi _2\rangle \dots |\chi _N\rangle \ .
\end{eqnarray}
The first and third terms will result in total derivatives that give zero when one integrates
over some variables \cite{1} and so what remain are just the second and third terms.

The commutator $P$ is given by a pure conformal transformation, and it is a function of the
generators ${\cal L}_n^E\ (n=-1,\dots )$ only. As the BRST charge commutes with all
${\cal L}_n^E$'s, i.e.
\begin{equation}
\left[ {\cal L}^E_n,Q\right] =0
\end{equation}
we have
\begin{equation}
\left[P,Q\right] =0\ .
\end{equation}
Considering now that
\begin{equation}
\left[ b_n^i,Q\right] ={\cal L}_n^i\ ,
\end{equation}
we then have, for $G$ given by (\ref{eq4.38}),
\begin{equation}
\label{eq4.39}
U_1\left[ G,Q\right] =U_1\sum_{i=1}^{N_1}\sum_{n=-1}^\infty \alpha _n^i{\cal L}_n^i\ .
\end{equation}

In order for the scattering amplitude $W$ to be BRST invariant, expression (\ref{eq4.39})
must be zero or a total derivative (that can be integrated out to become a null surface term).
At the same time, we want these extra ghosts to place (talking in terms of the simple cycling)
a ghost on one of the legs in $U_1$ that do not have any ghosts attached to them. If we now
remember property (\ref{eq4.27}), we see that we can satisfy these constraints in a nice way
by choosing $G$ to be given by
\begin{equation}
\label{eq4.40}
G=(-1)^{N_1+a}\sum_{j=1}^{N_1}\sum_{n=-1}^\infty e_n^{aj}b_n^j
\end{equation}
where $a\ (a\neq E)$ is one of the legs of vertex $U_1$ that does not have ghosts attached to
it. Inserting these ghosts in vertex $U_1$, we have
\begin{equation}
\label{eq4.41}
U_1G = V_1\prod_{\scriptstyle i=1\atop \scriptstyle i\neq a,b,c}^{N_1}\sum_{k=1}^{N_1}
\sum_{m=-1}^\infty e_n^{ik}b_m^k\times (-1)^{N_1+a}\sum_{j=1}^{N_1}\sum_{n=-1}^\infty e_n^{aj}
b_n^j = V_1\prod_{\scriptstyle i=1\atop \scriptstyle i\neq b,c}^{N_1}\sum_{j=1}^{N_1}
\sum_{n=-1}^\infty e_n^{ij}b_n^j\ .
\end{equation}

Using formula (\ref{eq4.40}) for the extra ghosts $G$, we then have
\begin{eqnarray}
U_1\left[ G,Q\right]   & = &  U_1(-1)^{N_1+a} \sum_{j=1}^{N_1} \sum_{n=-1}^\infty e_n^{aj}
\left[ b_n^j,Q\right] = (-1)^{N_1 +a}U_1\sum_{j=1}^{N_1} \sum_{n=-1}^\infty e_n^{aj}
{\cal L}_n^j \nonumber \\
 & = &  (-1)^{N_1+a} \frac{\partial V_1}{\partial z_a} \prod_{\scriptstyle i=1\atop
\scriptstyle i\neq a,b,c}^{N_1}\sum_{j=1}^{N_1} \sum_{n=-1}^\infty e_n^{ij}b_n^j
\end{eqnarray}
what is a total derivative that will vanish when one does the integration over $z_a$.

In references \cite{5} and \cite{6}, the extra ghosts have been placed in the propagator.
Although this can be done, there is no way one can derive a formula for the ghosts in the
propagator for a general cycling. In that case, the extra ghosts must be derived and BRST
invariance has to be checked for each particular cycling. Also, the resulting composite vertex
obtained in that case is not similar in its ghost structure to an ordinary tree vertex,
although it has the correct ghost number.

\subsection{Overlap identities}

We must now use the overlap identities to determine the propagator that satisfies them. In
order to do this we shall start with vertex $U_1^0$, which is the vertex with cycling
transformations $V_i^{-1}$ which involve leg $E$. Considering equations (\ref{eq4.28}) and
(\ref{eq4.29}), the overlap identities for vertex $U_1^0$ between an arbitrary leg $i$ and
leg $E$ are given by (figure 12)

\hskip 0.8 cm \includegraphics{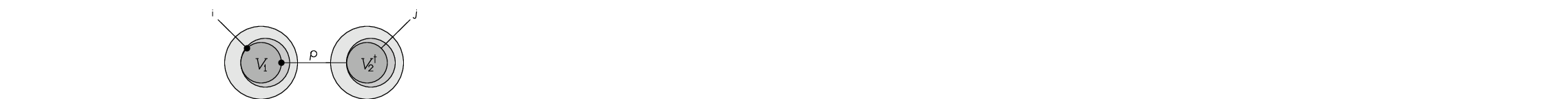}
\begin{center}
{\small Figure 12: Overlap identity for $U_1^0$.}
\end{center}

\begin{eqnarray}
\label{eq4.42}
 & & U_1^0\left[ b^i(\xi _{0i})-\left( \frac{d\xi _{0E}}{d\xi _{0i}}\right) ^2b^E(\xi _{0E})
\right] =0\ ,\\
 & & U_1^0\left[ c^i(\xi _{0i})-\left( \frac{d\xi _{0E}}{d\xi _{0i}}\right) ^{-1}c^E(\xi _{0E})
\right] \nonumber \\
\label{eq4.43}
 & & +V^0_1\sum_{\scriptstyle p=1\atop p\neq \scriptstyle a,b,c}^{N_1}(-1)^p
\prod _{\scriptstyle k=1\atop {\scriptstyle k\neq a,b,c\atop \scriptstyle k\neq p}}^{N_1}
\sum_{l=1}^{N_1}\sum_{q=-1}^\infty e_q^{kl}b_q^l \sum_{n=-1}^\infty \left[
e_n^{pi}(\xi _{0i})^{n+1}-\left( \frac{d\xi _{0E}}{d\xi _{0i}}\right) ^{-1}e_n^{pE}
(\xi _{0E})^{n+1} \right] =0\ .
\end{eqnarray}

The extra ghosts must then be inserted in vertex $U_1^0$ so that the composite vertex will
have the correct ghost number. Multiplying expressions (\ref{eq4.42}) and (\ref{eq4.43}) by
the extra ghosts $G$ (given by (\ref{eq4.40})) and passing them through the overlaps we
obtain (figure 13)

\hskip 0.8 cm \includegraphics{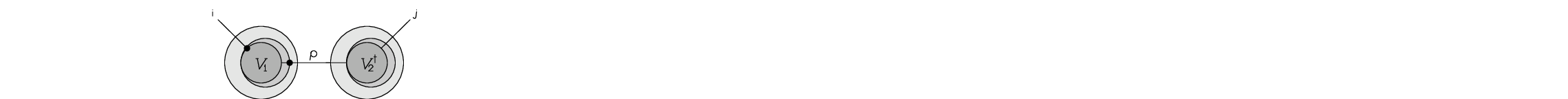}
\begin{center}
{\small Figure 13: Overlap identity for $U_1^0G$.}
\end{center}

\begin{eqnarray}
\label{eq4.44}
 & & U_1^0G\left[ b^i(\xi _{0i})-\left( \frac{d\xi _{0E}}{d\xi _{0i}}\right) ^2
b^E(\xi _{0E})\right] =0\ ,\\
 & & U_1^0G\left[ c^i(\xi _{0i})-\left( \frac{d\xi _{0E}}{d\xi _{0i}}\right) ^{-1}
c^E(\xi _{0E})\right] +U^0_1(-1)^{N_1+a}\sum_{n=-1}^\infty \left[ e_n^{ai}(\xi _{0i})^{n+1}
-\left( \frac{d\xi _{0E}}{d\xi _{0i}}\right) ^{-1}e_n^{aE}(\xi _{0E})^{n+1} \right]
\nonumber \\
 & & +V_1^0\sum_{\scriptstyle p=1\atop p\neq \scriptstyle a,b,c}^{N_1}(-1)^p
\prod _{\scriptstyle k=1\atop {\scriptstyle k\neq a,b,c\atop \scriptstyle k\neq p}}^{N_1}
\left( \sum_{l=1}^{N_1}\sum_{q=-1}^\infty e_q^{kl}b_q^l\right) \left( \sum_{j=1}^{N_1}
\sum_{n=-1}^\infty e_n^{aj}b_n^j\right) \nonumber \\
\label{eq4.45}
 & & \times \sum_{n=-1}^\infty \left[ e_n^{pi}(\xi _{0i})^{n+1}-\left(
\frac{d\xi _{0E}}{d\xi _{0i}}\right) ^{-1}e_n^{pE}(\xi _{0E})^{n+1} \right] =0\ .
\end{eqnarray}
The second and third terms of equation (\ref{eq4.45}) can be combined so that it becomes
\begin{eqnarray}
 & & U_1^0G\left[ c^i(\xi _{0i})-\left( \frac{d\xi _{0E}}{d\xi _{0i}}\right) ^{-1}
c^E(\xi _{0E})\right] \nonumber \\
\label{eq4.46}
 & & +V_1^0\sum_{\scriptstyle p=1\atop p\neq \scriptstyle b,c}^{N_1}(-1)^p
\prod _{\scriptstyle k=1\atop {\scriptstyle k\neq b,c\atop \scriptstyle k\neq p}}^{N_1}
\left( \sum_{l=1}^{N_1}\sum_{q=-1}^\infty e_q^{kl}b_q^l\right) \sum_{n=-1}^\infty
\left[ e_n^{pi}(\xi _{0i})^{n+1}-\left( \frac{d\xi _{0E}}{d\xi _{0i}}\right) ^{-1}
e_n^{pE}(\xi _{0E})^{n+1} \right] =0\ .
\end{eqnarray}

At this point, we must introduce conformal transformations of the type of $C_1$, given by
(\ref{eq4.11}) in order to have at the end the correct cycling transformations for the
composite vertex. In order to do this we need to use matrices $E_{nm}(\gamma )$, defined by
\cite{5}
\begin{equation}
\label{eq4.47}
E_{nm}(\gamma )=\left. \frac{1}{(m+1)!}\frac{\partial ^{m+1}}{\partial z^{m+1}}\left[
\left( \gamma z\right) ^{n+1}\left( \frac{\partial }{\partial z}\gamma z\right) ^{-1}\right]
\right| _{z=0}
\end{equation}
which have the following properties:
\begin{eqnarray}
\label{eq4.48}
 & & \sum_{t=-1}^1E_{rt}(\gamma _1)E_{ts}(\gamma _2)=E_{rs}(\gamma _1\gamma _2)\ \ ,\ \
r,s,t=-1,0,1\ ,\\
\label{eq4.49}
 & & E_{rn}(\gamma )=0\ \ ,\ \ r=-1,0,1\ \ ,\ \ n\geq 2\ ,\\
\label{eq4.50}
 & & \sum_{p=-1}^\infty E_{np}(\gamma _1)E_{pm}(\gamma _2)=E_{nm}(\gamma _1\gamma _2)\ \ ,\ \
n,m\geq -1\ ,\\
\label{eq4.51}
 & & \sum_{p=2}^\infty E_{np}(\gamma _1)E_{pm}(\gamma _2)=E_{nm}(\gamma _1\gamma _2)
-\sum_{r,s=-1}^1E_{nr}(\gamma _1)E_{rs}(\gamma _2)\delta _{sm}\ \ ,\ \ n,m\geq 2\ .
\end{eqnarray}

The action of the operator $C_1$ on the ghosts $b_n^i$ can then be calculated in the
following way: first we write $b_n^i$ in terms of an integral over the conformal operator
$b_n^i(\xi _i)$
\begin{equation}b_n^i=\oint _{\xi _{0i}=0}d\xi _{0i}\ (\xi _{0i})^{n+1}b^i(\xi _{0i})\ .
\end{equation}
Then we insert the operator $C_1$:
\begin{equation}
C_1b_n^iC_1^{-1}=\oint _{\xi _{0i}=0}d\xi _{0i}\ (\xi _{0i})^{n+1}\left( \frac{d}{d\xi _{0i}}
V_i^{-1}V_{0i}\xi _{0i}\right) ^2b^i\left( V_i^{-1}V_{0i}\xi _{0i}\right) \ .
\end{equation}
After a change of variables $\xi _i=V_i^{-1}V_{0i}\xi _{0i}$ we have\footnote{Note that,
because $\xi _i$ is a polynomial in $\xi _{0i}$ (with no constant term), then $\xi _{0i}=0
\Rightarrow \xi _i=0$.}
\begin{equation}
C_1b_n^iC_1^{-1}=\oint _{\xi _i=0}d\xi _i\ \left( V_{0i}^{-1}V_i\xi _i\right) ^{n+1}
\left( \frac{d}{d\xi _i}V_{0i}^{-1}V_i\xi _i\right) ^{-1}b^i(\xi _i) \ .
\end{equation}
Using matrices $E_{nm}(\gamma )$, we then may expand $\xi _{0i}$ in terms of $\xi _i$.
If we also expand $b^i(\xi _i)$, we then obtain
\begin{equation}
C_1b_n^iC_1^{-1}=\sum_{m=-1}^\infty \sum_{p=-\infty }^\infty \oint _{\xi _i=0}d\xi _i\
E_{nm}(V_{0i}^{-1}V_i)(\xi _i)^{m+1}b_{-p}^i(\xi _i)^{p-2}\ .
\end{equation}
Performing the integration we then have
\begin{equation}
\label{eq4.52}
C_1b_n^iC_1^{-1}=\sum_{m=-1}^\infty E_{nm}(V_{0i}^{-1}V_i)b_m^i\ .
\end{equation}

Using (\ref{eq4.52}) in equations (\ref{eq4.44}) and (\ref{eq4.46}) and multiplying
(\ref{eq4.44}) by $\left( d\xi _i/d\xi _{0i}\right) ^{-2}$ and (\ref{eq4.46}) by
$d\xi _i/d\xi _{0i}$, we then have (figure 14)

\includegraphics{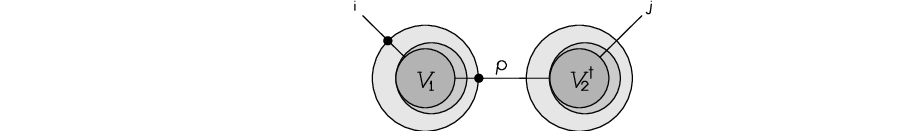}
\begin{center}
{\small Figure 14: Overlap identity for $U_1^0GC_1^{-1}$.}
\end{center}

\begin{eqnarray}
\label{eq4.55}
 & & U_1^0GC_1^{-1}\left[ b^i(\xi _i)-\left( \frac{d\xi _E}{d\xi _i}\right) ^2
b^E(\xi _E)\right] =0\ ,\\
 & & U_1^0GC_1^{-1} \left[ c^i(\xi _i) -\left( \frac{d\xi _E}{d\xi _i}\right) ^{-1}
c^E(\xi _E)\right] +V_1^0C_1^{-1} \sum_{\scriptstyle p=1\atop p\neq
\scriptstyle b,c}^{N_1}(-1)^p \prod _{\scriptstyle k=1\atop {\scriptstyle k\neq b,c
\atop \scriptstyle k\neq p}}^{N_1} \left[ \sum_{l=1}^{N_1} \sum_{q,t=-1}^\infty e_q^{kl}
E_{qt}\left( V_{0l}^{-1}V_l\right) b_t^l\right] \nonumber \\
\label{eq4.56}
 & & \times \sum_{n,m=-1}^\infty \left[ e_n^{pi}E_{nm}\left( V_{0i}^{-1}V_i\right)
(\xi _i)^{m+1}  -\left( \frac{d\xi _E}{d\xi _i}\right) ^{-1}e_n^{pE}E_{nm}\left( V_{0E}^{-1}
V_E\right) (\xi _E)^{m+1} \right] =0\ .
\end{eqnarray}

Before going further, some words must be said about the effects of $C_1^{-1}$ on vertex
$U_1^0$ with the extra ghosts $G$. This is given explicitly by \cite{4}\cite{5}
\begin{eqnarray}
U_1^0G & = &  \left( \prod_{i=1}^{N_1} { }_i\langle 0|\right)  \exp \left[
\sum_{\scriptstyle i,j=1 \atop \scriptstyle i\neq j}^{N_1} \sum_{n=2}^\infty
\sum_{m=-1}^\infty c_n^i E_{nm} \left( \Gamma V^{-1}_{0i}V_{0j}\right) b_m^j\right]
\nonumber \\
 & & \times \prod_{r=-1}^1 \sum_{i=1}^{N_1} \sum_{s=-1}^1 E_{rs}(V_{0i})b_s^i \times
\prod_{\scriptstyle i=1\atop \scriptstyle i\neq b,c}^{N_1} \sum_{j=1}^N \sum_{n=-1}^\infty
e_n^{ij}b_n^j\ .
\end{eqnarray}

Making use of matrices $F_{nm}(\gamma )$, defined by \cite{5}
\begin{equation}
\label{eq4.57}
F_{nm}(\gamma )=\left. \frac{1}{(m-2)!}\frac{\partial ^{m-2}}{\partial z^{m-2}}\left\{
\left[ \gamma (z)\right] ^{n-2}\left[ \frac{\partial }{\partial z}\gamma (z)\right] ^{-1}
\right\} \right| _{z=0}
\end{equation}
we may calculate in a similar way as we did for the $b_n^i$ ghosts the effect of $C_1$ on
the $c_n^i$ ghosts, obtaining
\begin{equation}C_1c_n^iC_1^{-1}=\sum_{m=2}^\infty F_{nm}(V_{0i}^{-1}V_i)c_m^i\ .
\end{equation}
Using this together with the property
\begin{equation}F_{nm}(\gamma )=E_{mn}(\Gamma \gamma ^{-1}\Gamma )\end{equation}
and equation (\ref{eq4.52}), we may then show that the result of acting with $C_1^{-1}$
on $U_1^0G$ is
\begin{eqnarray}
U_1^0GC_1^{-1} & = &  \left( \prod_{i=1}^{N_1} { }_i\langle 0|\right) \exp \left[
\sum_{\scriptstyle i,j=1 \atop \scriptstyle i\neq j}^{N_1} \sum_{n=2}^\infty
\sum_{m=-1}^\infty c_n^i E_{nm} \left( \Gamma V^{-1}_iV_j\right) b_m^j\right]
\nonumber \\
\label{eq4.58}
 & & \times \prod_{r=-1}^1 \sum_{i=1}^{N_1} \sum_{s=-1}^1 E_{rs}(V_i)b_s^i \times
\prod_{\scriptstyle i=1 \atop \scriptstyle i\neq b,c}^{N_1} \sum_{j=1}^N
\sum_{n,m=-1}^\infty e_n^{ij}E_{nm}(V_{0j}V_j)b_m^j\ .
\end{eqnarray}
So we can see that in this case the action of $C_1^{-1}$ on $U_1^0G$ is not just to
change $V_{0i}\rightarrow V_i$. Because of the peculiar nature of $e_n^{ij}$, it transforms
as
\begin{equation}
e_n^{ij}\rightarrow \sum_{m=-1}^\infty e_n^{ij}E_{nm}(V_{0j}V_j)b_m^j\ .
\end{equation}
Only in one particular group of cycling transformations (as we shall see later) will this be
just equivalent to changing $V_{0i}\rightarrow V_i$. We shall call from now on $U_1^0GC_1^{-1}
\equiv U_1$ and $V_1^0C_1^{-1}\equiv V_1$. The calculation for vertex $C_FU_2^\dagger C_2^{-1}$
will be similar to the one we have just made for $U_1^0GC_1^{-1}$.

Having done this, we must insert the propagator ${\cal P}$ into the overlap identities
(\ref{eq4.55}) and (\ref{eq4.56}) in the same way as in the case with no ghosts. But now we
must take extra care since there are terms depending on $b_q^E$ in the second term of
equation (\ref{eq4.56}). Using equation (\ref{eq4.52}) as a guideline, we have
\begin{equation}{\cal P}^{-1}b_q^E{\cal P}=\sum_{t=-1}^\infty E_{qt}({\cal P})b_t^E
\end{equation}
so that the result of inserting ${\cal P}$ into overlaps (\ref{eq4.55}) and (\ref{eq4.56})
is (figure 15)

\hskip 0.8 cm \includegraphics{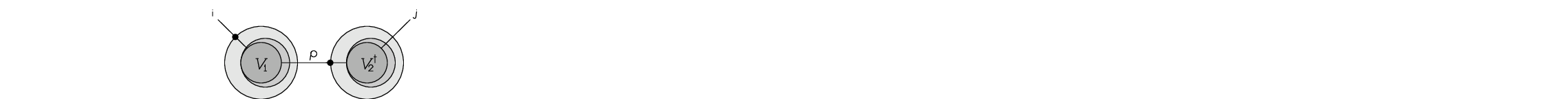}
\begin{center}
{\small Figure 15: Overlap identity for $U_1{\cal P}$.}
\end{center}

\begin{eqnarray}
\label{eq4.59}
 & & U_1{\cal P}\left[ b^i(\xi _i)-\left( \frac{d}{d\xi _i}{\cal P}^{-1}\xi _E\right) ^2b^E
\left( {\cal P}^{-1}\xi _E\right) \right] =0\ ,\\
 & & U_1{\cal P}\left[ c^i(\xi _i)-\left( \frac{d}{d\xi _i}{\cal P}^{-1}\xi _E\right) ^{-1}c^E
\left( {\cal P}^{-1}\xi _E\right) \right] \nonumber \\
 & & +V_1{\cal P}\sum_{\scriptstyle p=1\atop p\neq \scriptstyle b,c}^{N_1}(-1)^p
\prod _{\scriptstyle k=1\atop {\scriptstyle k\neq b,c\atop \scriptstyle k\neq p}}^{N_1}
\sum_{q,t=-1}^\infty \left[ \sum_{\scriptstyle l=1\atop \scriptstyle l\neq E}^{N_1} e_q^{kl}
E_{qt}(V_{0l}^{-1}V_l)b_t^l +e_q^{kE}E_{qt}(V_{0E}^{-1}V_E{\cal P})b_t^E\right] \nonumber \\
\label{eq4.60}
 & & \times \sum_{n,m=-1}^\infty \left[ e_n^{pi}E_{nm}(V_{0i}^{-1}V_i)(\xi _i)^{m+1}
-\left( \frac{d\xi _E}{d\xi _i}\right) ^{-1}e_n^{pE}E_{nm}(V_{0E}^{-1}V_E)(\xi _E)^{m+1}
\right] =0\ .
\end{eqnarray}

The operators of the overlap equations are now facing leg $F$ of vertex $U_2^\dagger $. In
order to obtain the overlap identities for this leg, we must now identify the operators of
leg $E$ with the ones of leg $F$, which are adjoint operators:
\begin{equation}b^E_t\rightarrow {b^F_t}^\dagger \ \ ,\ \ b^E\rightarrow {b^F}^\dagger \ \ ,
\ \ c^E\rightarrow {c^F}^\dagger\ .\end{equation}
First, as $b^F$ and $c^F$ are conformal operators with weights 2 and $-1$, respectively, we
have
\begin{eqnarray}
{b^F}^\dagger (\xi _F) & = & \Gamma b^F(\xi _F)\Gamma =\left( \frac{d}{d\xi _F}\Gamma
\xi _F\right) ^2b^F\left( \Gamma \xi _F\right) =(\xi _F)^{-4}b^F(\Gamma \xi _F)\ ,\\
{c^F}^\dagger (\xi _F) & = & \Gamma c^F(\xi _F)\Gamma =\left( \frac{d}{d\xi _F}\Gamma
\xi _F\right) ^{-1}c^F\left( \Gamma \xi _F\right) -(\xi _F)^2c^F(\Gamma \xi _F)\ .
\end{eqnarray}
Then, for $b_n^F$ and $c_n^F$, we obtain
\begin{eqnarray}
{c_n^F}^\dagger & = &\Gamma c_n^F\Gamma = -c_{-n}^F\ ,\\
{b_n^F}^\dagger & = & \Gamma b_n^F\Gamma b_{-n}^F\ .
\end{eqnarray}
Then, we must also make the change
\begin{eqnarray}
 \left( \frac{d\xi _E}{d\xi _i}\right) ^{-1}e_n^{pE}(\xi _E)^{n+1}  & = &
\left( \frac{d\xi _F}{d\xi _i}\right) ^{-1}\left( \frac{d\xi _E}{d\xi _F}\right) ^{-1}
e_n^{pE}(\xi _E)^{n+1} \nonumber \\
 & = &  \left( \frac{d\xi _F}{d\xi _i}\right) ^{-1}\sum_{m=-1}^\infty e_n^{pE}E_{nm}
\left( V_E^{-1}V_F\right) (\xi _F)^{m+1} \ .
\end{eqnarray}

So, the overlap equations become
\begin{eqnarray}
\label{eq4.61}
 & & U_1{\cal P}\left[ b^i(\xi _i)-\left( \frac{d}{d\xi _i}{\cal P}^{-1}\xi _E\right) ^2
\left( {\cal P}^{-1}\xi _E\right) ^{-4}b^F\left( \Gamma {\cal P}^{-1}\xi _E\right) \right] =0
\ ,\\
 & & U_1{\cal P}\left[ c^i(\xi _i)+\left( \frac{d}{d\xi _i}{\cal P}^{-1}\xi _E\right) ^{-1}
\left( {\cal P}^{-1}\xi _E\right) ^2c^F\left( \Gamma {\cal P}^{-1}\xi _E\right) \right]
\nonumber \\
 & & +V_1{\cal P}\sum_{\scriptstyle p=1\atop p\neq \scriptstyle b,c}^{N_1}(-1)^p
\prod _{\scriptstyle k=1\atop {\scriptstyle k\neq b,c\atop \scriptstyle k\neq p}}^{N_1}
\sum_{q,t=-1}^\infty \left[ \sum_{\scriptstyle l=1\atop \scriptstyle l\neq E}^{N_1}e_q^{kl}
E_{qt}\left( V_{0l}^{-1}V_l\right) b_t^l +e_q^{kE}E_{qt}\left( V_{0E}^{-1}V_E{\cal P}\right)
b_{-t}^F\right] \nonumber \\
\label{eq4.62}
 & & \times \sum_{n,m=-1}^\infty \left[ e_n^{pi}E_{nm}\left( V_{0i}^{-1}V_i\right)
(\xi _i)^{m+1}  -\left( \frac{d\xi _F}{d\xi _i}\right) ^{-1}e_n^{pE}E_{nm}\left( V_{0E}^{-1}
V_F\right) (\xi _F)^{m+1}\right] =0\ .
\end{eqnarray}

We are then facing the conformal transformation $C_F$ that takes $\xi _F$ into $\xi _{0F}$.
Inserting this transformation we obtain (figure 16):

\hskip 0.8 cm \includegraphics{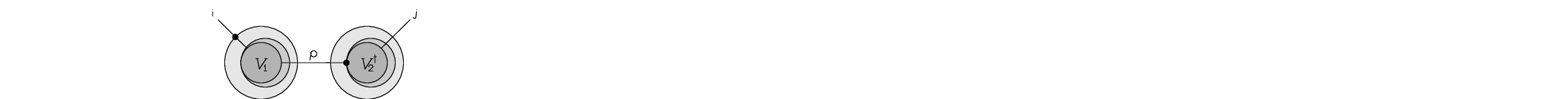}
\begin{center}
{\small Figure 16: Overlap identity for $U_1{\cal P}C_F$.}
\end{center}

\begin{eqnarray}
 & & U_1{\cal P}C_F \left[ b^i(\xi _i)- \left( \frac{d}{d\xi _i}{\cal P}^{-1}\xi _E\right) ^2
\left( {\cal P}^{-1}\xi _E\right) ^{-4} \right. \nonumber \\
\label{eq4.63}
 & & \ \ \ \ \ \ \ \ \ \ \ \ \ \ \ \ \ \ \ \ \ \ \ \ \ \ \ \ \ \ \ \ \ \ \ \ \ \ \ \ \ \
\times \left. \left( \frac{dV_{0F}^{-1}V_F\Gamma {\cal P}^{-1}\xi _E}{d\Gamma {\cal P}^{-1}
\xi _E}\right) ^2 b^F \left( V_{0F}^{-1}V_F\Gamma {\cal P}^{-1}\xi _E\right) \right] =0 \ ,\\
 & & U_1{\cal P}C_F\left[ c^i(\xi _i)+\left( \frac{d}{d\xi _i}{\cal P}^{-1}\xi _E\right) ^{-1}
\left( {\cal P}^{-1}\xi _E\right) ^2 \left( \frac{dV_{0F}^{-1}V_F\Gamma {\cal P}^{-1}\xi _E}
{d\Gamma {\cal P}^{-1}\xi _E}\right) ^{-1}c^F\left( V_{0F}^{-1}V_F\Gamma {\cal P}^{-1}\xi _E
\right) \right]\nonumber \\
 & & +V_1{\cal P}C_F\sum_{\scriptstyle p=1\atop p\neq \scriptstyle b,c}^{N_1}(-1)^p
\prod _{\scriptstyle k=1\atop {\scriptstyle k\neq b,c\atop \scriptstyle k\neq p}}^{N_1}
\sum_{q,t=-1}^\infty \left[ \sum_{\scriptstyle l=1\atop \scriptstyle l\neq E}^{N_1}e_q^{kl}
E_{qt}\left( V_{0l}^{-1}V_l\right) b_t^l -e_q^{kE}E_{qt}\left( V_{0E}^{-1}V_E{\cal P}\Gamma
V_F^{-1}V_{0F}\right) b_t^F\right] \nonumber \\
\label{eq4.64}
 & & \times \sum_{n,m=-1}^\infty \left[ e_n^{pi}E_{nm}\left( V_{0i}^{-1}V_i\right)
(\xi _i)^{m+1} -\left( \frac{d\xi _{0F}}{d\xi _i}\right) ^{-1}e_n^{pE}E_{nm}\left( V_{0E}^{-1}
V_{0F}\right) (\xi _{0F})^{m+1}\right] =0\ .
\end{eqnarray}
We are facing now vertex $V_2^{0\dagger }$. This vertex satisfies the following overlap
identity \cite{4}:
\begin{equation}
\label{eq4.65}
\sum_{s=-1}^1E_{rs}(V_{0F})b_{-s}^FV_2^{0\dagger }=-V_2^{0\dagger }\sum_{\scriptstyle i=1
\atop \scriptstyle i\neq F}^{N_2}\sum_{s=-1}^1E_{rs}(V_{0i})b_s^i\ \ ,\ \ r=-1,0,1\ .
\end{equation}

Using equation (\ref{eq4.50}), we have
\begin{equation}E_{qt}(V_{0E}^{-1}V_E{\cal P}\Gamma V_F^{-1}V_{0F})b_t^F=\sum_{u=-1}^\infty
E_{qu}(V_{0E}^{-1}V_E{\cal P}\Gamma V_F^{-1})E_{ut}(V_{0F})b_t^F\ .\end{equation}
All terms on $b_t^F$ with $t\geq 2$ get annihilated by the conjugate vacuum $|0\rangle _F$,
while we may use identity (\ref{eq4.65}) to substitute the terms in $b_r^F$, $r=-1,0,1$. Doing
this, equation (\ref{eq4.64}) becomes
\begin{eqnarray}
 & & U_1{\cal P}C_F\left[ c^i(\xi _i)+\left( \frac{d}{d\xi _i}{\cal P}^{-1}\xi _E\right) ^{-1}
\left( {\cal P}^{-1}\xi _E\right) ^2\left( \frac{dV_{0F}^{-1}V_F\Gamma {\cal P}^{-1}\xi _E}
{d\Gamma {\cal P}^{-1}\xi _E}\right) ^{-1}c^F\left( V_{0F}^{-1}V_F\Gamma {\cal P}^{-1}\xi _E
\right) \right] \nonumber \\
 & & +V_1{\cal P}C_F\sum_{\scriptstyle p=1\atop p\neq \scriptstyle b,c}^{N_1}(-1)^p
\prod _{\scriptstyle k=1\atop {\scriptstyle k\neq b,c\atop \scriptstyle k\neq p}}^{N_1}
\sum_{q,t=-1}^\infty \left[ \sum_{\scriptstyle l=1\atop \scriptstyle l\neq E}^{N_1}e_q^{kl}
E_{qt}\left( V_{0l}^{-1}V_l\right) b_t^l+\sum_{\scriptstyle l=1\atop \scriptstyle l\neq F}^{N_2}
e_q^{kE}E_{qt}\left( V_{0E}^{-1}V_E{\cal P}\Gamma V_F^{-1}V_{0l}\right) b_t^l\right]
\nonumber \\
\label{eq4.66}
 & & \times \sum_{n,m=-1}^\infty \left[ e_n^{pi}E_{nm}\left( V_{0i}^{-1}V_i\right)
(\xi _i)^{m+1}  -\left( \frac{d\xi _{0F}}{d\xi _i}\right) ^{-1}e_n^{pE}E_{nm}\left(
V_{0E}^{-1}V_{0F}\right) (\xi _{0F})^{m+1}\right] =0\ .
\end{eqnarray}
Equations (\ref{eq4.63}) and (\ref{eq4.66}) are the overlap identities between legs $i$ and
$F$.

In order to obtain the overlaps between leg $i$ of vertex $U_1$ and an arbitrary leg $j$ of
vertex $U_2^\dagger $, we must now perform a cycling transformation that will take the
operators from leg $F$ to leg $j$. The effect of this transformation on $c^F\left( \Gamma
{\cal P}^{-1}V_E^{-1}V_i\xi _i\right) $ is
\begin{equation}
 V_j^{-1}V_Fc^F\left( V_{0F}^{-1}V_F\Gamma {\cal P}^{-1}\xi _E\right) V_F^{-1}V_j=\left(
\frac{dV_{0j}^{-1}V_F\Gamma {\cal P}^{-1}\xi _E}{dV_{0F}^{-1}V_F\Gamma {\cal P}^{-1}\xi _E}
\right) ^{-1}c^j\left( V^{-1}_jV_F\Gamma {\cal P}^{-1}\xi _E\right) \ .
\end{equation}
Then, we must also write
\begin{equation}
\left( \frac{d\xi _{0F}}{d\xi _i}\right) ^{-1}\sum_{m=-1}^\infty e_n^{pE}E_{nm}\left(
V_{0E}^{-1}V_{0F}\right) (\xi _{0F})^{m+1} =\left( \frac{d\xi _{0j}}{d\xi _i}\right) ^{-1}
\sum_{m=-1}^\infty e_n^{pE}E_{nm}\left( V_{0E}^{-1}V_{0j}\right) (\xi _{0j})^{m+1}\ .
\end{equation}
Doing this, overlap equations (\ref{eq4.63}, \ref{eq4.66}) become (figure 17)

\hskip 0.8 cm \includegraphics{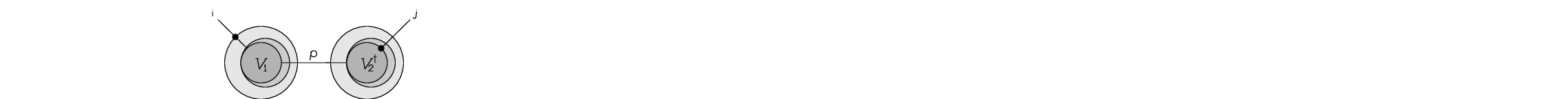}
\begin{center}
{\small Figure 17: Overlap identity for $U_1{\cal P} C_FV_2^{0\dagger }$.}
\end{center}

\begin{eqnarray}
 & & U_1{\cal P} C_FV_2^{0\dagger }\left[ b^i(\xi _i)-\!  \left( \frac{d}{d\xi _i}
{\cal P}^{-1}\xi _E\right) ^2\left( {\cal P}^{-1}\xi _E\right) ^{-4} \right. \nonumber \\
\label{eq4.67}
 & &  \ \ \ \ \ \ \ \ \ \ \ \ \ \ \ \ \ \ \ \ \ \ \ \ \ \ \ \ \ \ \ \ \ \ \ \ \ \ \ \ \ \
\left. \times \left( \frac{dV_{0j}^{-1} V_F\Gamma {\cal P}^{-1}\xi _E} {d\Gamma {\cal P}^{-1}
\xi _E}\right) ^2 b^F \left( V_{0j}^{-1}V_F \Gamma {\cal P}^{-1} \xi _E\right) \right] =0\ ,\\
 & &  U_1{\cal P}C_FV_2^{0\dagger }\left[ c^i(\xi _i)+\left( \frac{d}{d\xi _i}{\cal P}^{-1}
\xi _E\right) ^{-1}\left( {\cal P}^{-1}\xi _E\right) ^2 \left( \frac{dV_{0j}^{-1}V_F\Gamma
{\cal P}^{-1}\xi _E}{d\Gamma {\cal P}^{-1}\xi _E}\right) ^{-1}c^F\left( V_{0j}^{-1}V_F\Gamma
{\cal P}^{-1}\xi _E\right) \right] \nonumber \\
 & & +V_1{\cal P}C_FV_2^{0\dagger }\! \sum_{\scriptstyle p=1\atop p\neq \scriptstyle b,c}^{N_1}
(-1)^p\! \! \! \! \prod _{\scriptstyle k=1\atop {\scriptstyle k\neq b,c\atop \scriptstyle k\neq
p}}^{N_1} \! \sum_{q,t=-1}^\infty \left[ \sum_{\scriptstyle l=1\atop \scriptstyle l
\neq E}^{N_1}e_q^{kl}E_{qt}\left( V_{0l}^{-1}V_l\right) b_t^l +\sum_{\scriptstyle l=1\atop
\scriptstyle l\neq F}^{N_2}e_q^{kE}E_{qt}\left( V_{0E}^{-1}V_E{\cal P}\Gamma V_F^{-1}V_{0l}
\right) b_t^l\right] \nonumber \\
\label{eq4.68}
 & & \times \sum_{n,m=-1}^\infty \left[ e_n^{pi}E_{nm}\left( V_{0i}^{-1}V_i\right)
(\xi _i)^{m+1}  -\left( \frac{d\xi _{0j}}{d\xi _i}\right) ^{-1}e_n^{pE}E_{nm}\left(
V_{0E}^{-1}V_{0j}\right) (\xi _{0j})^{m+1}\right] =0\ .
\end{eqnarray}

The operators are now facing the ghosts that surround vertex $U_2^{0\dagger }$ (like in
equation (\ref{eq4.26})):
\begin{equation}\prod_{\scriptstyle k=1\atop \scriptstyle k\neq d,g,h}^{N_2}\left(
\sum_{\scriptstyle l=1\atop \scriptstyle l\neq F}^{N_2}\sum_{q=-1}^\infty e_q^{kl}b_q^l
+\sum_{q=-1}^\infty e_q^{kF}b_{-q}^F\right) \ .\end{equation}
so that we must insert these ghosts into the expressions for the overlaps. Before doing that,
we must notice that the extra ghosts acting on vertex $V_2^{0\dagger }$ have at their left both
the conformal transformation $C_F$ and the propagator ${\cal P}$ so that we must first pass
them through in order to reach vertex $U_1$:
\begin{eqnarray}
 & & {\cal P}C_F\prod_{\scriptstyle k=1\atop \scriptstyle k\neq d,g,h}^{N_2}\left(
\sum_{\scriptstyle l=1\atop \scriptstyle l\neq F}^{N_2}\sum_{q=-1}^\infty e_q^{kl}b_q^l
+\sum_{q=-1}^\infty e_q^{kF}b_{-q}^F\right) \nonumber \\
 & & \ \ \ \ =\prod_{\scriptstyle k=1\atop \scriptstyle k\neq d,g,h}^{N_2}\left(
\sum_{\scriptstyle l=1\atop \scriptstyle l\neq F}^{N_2}\sum_{q=-1}^\infty e_q^{kl}b_q^l
-\sum_{q,t=-1}^\infty e_q^{kF}E_{qt}(V_{0F}^{-1}V_F\Gamma {\cal P}^{-1})b_t^F\right)
{\cal P}C_F\ .
\end{eqnarray}
Then, identifying legs $E$ and $F$, we have the following expression for the ghosts:
\begin{equation}\prod_{\scriptstyle k=1\atop \scriptstyle k\neq d,g,h}^{N_2}\left[
\sum_{\scriptstyle l=1\atop \scriptstyle l\neq F}^{N_2}\sum_{q=-1}^\infty e_q^{kl}b_q^l
-\sum_{q,t=-1}^\infty e_q^{kF}E_{qt}\left( V_{0F}^{-1}V_F\Gamma {\cal P}^{-1}\right)
b_t^E\right] {\cal P}\ .\end{equation}
We may now pass it through the conformal transformation $C_1$, obtaining
\begin{equation}\prod_{\scriptstyle k=1\atop \scriptstyle k\neq d,g,h}^{N_2}\left[
\sum_{\scriptstyle l=1\atop \scriptstyle l\neq F}^{N_2}\sum_{q=-1}^\infty e_q^{kl}b_q^l
-\sum_{q,t=-1}^\infty e_q^{kF}E_{qt}\left( V_{0F}^{-1}V_F\Gamma {\cal P}^{-1}V_E^{-1}V_{0E}
\right) b_t^E\right] {\cal P}\ .\end{equation}

This is now facing vertex $V_1^0$, which satisfies the following overlap identity \cite{4}:
\begin{equation}V_1^0\sum_{s=-1}^1E_{rs}(V_{0E})b_s^E=-V_1^0\sum_{\scriptstyle i=1\atop
\scriptstyle i\neq E}^{N_1}\sum_{s=-1}^1E_{rs}(V_{0i})b_s^i\ \ ,\ \ r=-1,0,1\ .\end{equation}
Using this identity, we may then write the extra ghosts as
\begin{equation}\prod_{\scriptstyle k=1\atop \scriptstyle k\neq d,g,h}^{N_2}\left[
\sum_{\scriptstyle l=1\atop \scriptstyle l\neq F}^{N_2}\sum_{q=-1}^\infty e_q^{kl}b_q^l+
\sum_{\scriptstyle l=1\atop \scriptstyle l\neq E}^{N_1}\sum_{q,t=-1}^\infty e_q^{kF}E_{qt}
\left( V_{0F}^{-1}V_F\Gamma {\cal P}^{-1}V_E^{-1}V_{0l}\right) b_t^l\right] .\end{equation}
Passing it back through $C_1$ we then have
\begin{equation}
\label{eq4.69}
\prod_{\scriptstyle k=1\atop \scriptstyle k\neq d,g,h}^{N_2}\left[ \sum_{\scriptstyle l=1\atop
\scriptstyle l\neq F}^{N_2}\sum_{q=-1}^\infty e_q^{kl}b_q^l+\sum_{\scriptstyle l=1\atop
\scriptstyle l\neq E}^{N_1}\sum_{q,t=-1}^\infty e_q^{kF}E_{qt}\left( V_{0F}^{-1}V_F\Gamma
{\cal P}^{-1}V_E^{-1}V_l\right) b_t^l\right] .
\end{equation}

We now insert these ghosts into expressions (\ref{eq4.67}) and (\ref{eq4.68}). We do so by
multiplying them by (\ref{eq4.69}) and passing it through the first term of the overlaps.
What we obtain is (figure 18)

\newpage 

\hskip 0.8 cm \includegraphics{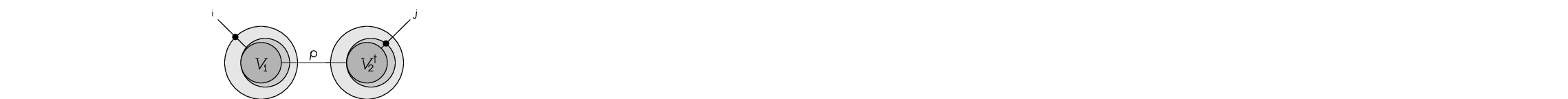}
\begin{center}
{\small Figure 18: Overlap identity for $U_1{\cal P}C_FU_2^{0\dagger }$.}
\end{center}

\begin{eqnarray}
 & & U_1{\cal P}C_FU_2^{0\dagger }\left[ b^i(\xi _i)-\left( \frac{d}{d\xi _i}{\cal P}^{-1}
\xi _E\right) ^2\left( {\cal P}^{-1}\xi _E\right) ^{-4}\right. \nonumber \\
\label{eq4.70}
 & & \ \ \ \ \ \ \ \ \ \ \ \ \ \ \ \ \ \ \ \ \ \ \ \ \ \ \ \ \ \ \ \ \ \ \ \ \ \ \ \ \ \
\left. \times \left( \frac{dV_{0j}^{-1}V_F\Gamma {\cal P}^{-1}\xi _E}{d\Gamma {\cal P}^{-1}
\xi _E}\right) ^2 b^F\left( V_{0j}^{-1}V_F\Gamma {\cal P}^{-1}\xi _E\right) \right] =0\ ,\\
 & & U_1{\cal P}C_FU_2^{0\dagger }\left[ c^i(\xi _i)+\left( \frac{d}{d\xi _i}{\cal P}^{-1}
\xi _E\right) ^{-1}\left( {\cal P}^{-1}\xi _E\right) ^2 \left( \frac{dV_{0j}^{-1}V_F\Gamma
{\cal P}^{-1}\xi _E}{d\Gamma {\cal P}^{-1}\xi _E}\right) ^{-1}c^F\left( V_{0j}^{-1}V_F\Gamma
{\cal P}^{-1}\xi _E\right) \right] \nonumber \\
 & & +U_1{\cal P}C_FV_2^{0\dagger }\sum_{p=1}^{N_2}(-1)^p\prod_{\scriptstyle k=1\atop
{\scriptstyle k\neq d,g,h\atop \scriptstyle k\neq p}}^{N_2}\left[ \sum_{\scriptstyle l=1\atop
\scriptstyle l\neq F}^{N_2}\sum_{q=-1}^\infty e_q^{kl}b_q^l +\sum_{\scriptstyle l=1\atop
\scriptstyle l\neq E}^{N_1}\sum_{q,t=-1}^\infty e_q^{kF}E_{qt}\left( V_{0F}^{-1}V_F\Gamma
{\cal P}^{-1}V_E^{-1}V_l\right) b_t^l\right] \nonumber \\
 & & \times \sum_{n=-1}^\infty \left[ \sum_{m=-1}^\infty e_n^{pF}E_{nm}\left( V_{0F}^{-1}V_F
\Gamma {\cal P}^{-1}V_E^{-1}V_i\right) (\xi _i)^{m+1} \right. \nonumber \\
 & & \ \ \ \ \ \ \ \ \ \ \ \ \ \ \ \  \left. +\left( \frac{d}{d\xi _i}{\cal P}^{-1}
\xi _E\right) ^{-1}\left( {\cal P}^{-1}\xi _E\right) ^2 \left( \frac{dV_{0j}^{-1}V_F
\Gamma {\cal P}^{-1}\xi _E}{d\Gamma {\cal P}^{-1}\xi _E}\right) ^2e_n^{pj}\left(
V_{0j}^{-1}V_F\Gamma {\cal P}^{-1}\xi _E\right) ^{n+1}\right] \nonumber \\
 & & +V_1{\cal P}C_FV_2^{0\dagger }\sum_{\scriptstyle p=1\atop \scriptstyle p\neq b,c}^{N_1}
(-1)^p\! \! \! \! \prod _{\scriptstyle k=1\atop {\scriptstyle k\neq b,c\atop \scriptstyle k
\neq p}}^{N_1}\sum_{q,t=-1}^\infty \left[ \sum_{\scriptstyle l=1\atop \scriptstyle l
\neq E}^{N_1}e_q^{kl}E_{qt}(V_{0l}^{-1}V_l)b_t^l +\sum_{\scriptstyle l=1\atop \scriptstyle l
\neq F}^{N_2}e_q^{kE}E_{qt}\left( V_{0E}^{-1}V_E{\cal P}\Gamma V_F^{-1}V_{0l}\right) b_t^l
\right] \nonumber \\
 & & \times \sum_{n,m=-1}^\infty \left[ e_n^{pi}E_{nm}(V_{0i}^{-1}V_i)(\xi _i)^{m+1}-\left(
\frac{d\xi _{0j}}{d\xi _i}\right) ^{-1}e_n^{pE}E_{nm}(V_{0E}^{-1}V_E)(\xi _{0j})^{m+1}\right]
\nonumber \\
\label{eq4.71}
 & & \times \prod_{\scriptstyle k=1\atop \scriptstyle k\neq d,g,h}^{N_2}\left[
\sum_{\scriptstyle l=1\atop \scriptstyle l\neq F}^{N_2}\sum_{q=-1}^\infty e_q^{kl}b_q^l
+\sum_{\scriptstyle l=1\atop \scriptstyle l\neq E}^{N_1}\sum_{q,t=-1}^\infty e_q^{kF}E_{qt}
\left( V_{0F}^{-1}V_F\Gamma {\cal P}^{-1}V_E^{-1}V_l\right) b_t^l\right] =0\ .
\end{eqnarray}

We are then facing the last term of this composite vertex: the conformal transformation
$C_2$ on vertex $V_2^{0\dagger }$. Inserting it into equations (\ref{eq4.70}) and
(\ref{eq4.71}), we obtain (figure 19)

\hskip 0.8 cm \includegraphics{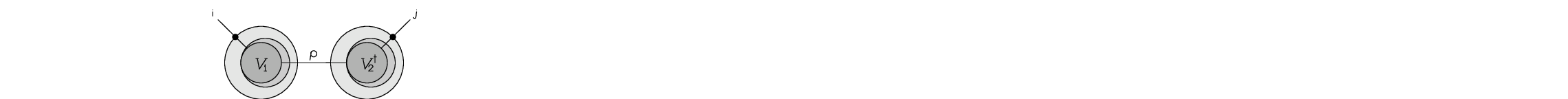}
\begin{center}
{\small Figure 19: Overlap identity for $U_1{\cal P}U_2^\dagger $.}
\end{center}

\begin{eqnarray}
 & & U_1{\cal P}U_2^\dagger \left[ b^i(\xi _i)-\left( \frac{d}{d\xi _i}{\cal P}^{-1}
\xi _E\right) ^2\left( {\cal P}^{-1}\xi _E\right) ^{-4} \right. \nonumber \\
\label{eq4.72}
 & & \ \ \ \ \ \ \ \ \ \ \ \ \ \ \ \ \ \ \ \ \ \ \ \ \ \ \ \ \ \ \ \ \ \ \ \ \ \ \ \ \ \
\left. \times \left( \frac{dV_j^{-1}V_F\Gamma {\cal P}^{-1}\xi _E}{d\Gamma {\cal P}^{-1}\xi _E}
\right) ^2 b^F\left( V_j^{-1}V_F \Gamma {\cal P}^{-1}\xi _E\right) \right] =0\ ,\\
 & & U_1{\cal P} U_2^\dagger \left[ c^i(\xi _i) +\left( \frac{d}{d\xi _i}{\cal P}^{-1} \xi _E
\right) ^{-1} \left( {\cal P}^{-1}\xi _E\right) ^2 \left( \frac{dV_j^{-1}  V_F\Gamma
{\cal P}^{-1} \xi _E}{d \Gamma {\cal P}^{-1} \xi _E}\right) ^{-1} c^F\left( V_j^{-1} V_F
\Gamma {\cal P}^{-1} \xi _E\right) \right] \nonumber \\
 & & +U_1 {\cal P} V_2^\dagger \sum_{p=1}^{N_2}(-1)^p \prod_{\scriptstyle k=1 \atop
{\scriptstyle k\neq d,g,h \atop \scriptstyle k\neq p}}^{N_2} \sum_{q,t=-1}^\infty \left[
\sum_{\scriptstyle l=1 \atop \scriptstyle l\neq F}^{N_2} e_q^{kl} E_{qt}(V_{0l}^{-1}V_l)
b_t^l+ \sum_{\scriptstyle l=1\atop \scriptstyle l\neq E}^{N_1} e_q^{kF}E_{qt}\left(
V_{0F}^{-1} V_F \Gamma {\cal P}^{-1} V_E^{-1}V_l\right) b_t^l\right] \nonumber \\
 & & \times \sum_{n=-1}^\infty  \left[ \sum_{m=-1}^\infty e_n^{pF}E_{nm} \left( V_{0F}^{-1}
V_F \Gamma {\cal P}^{-1} V_E^{-1} V_i\right) (\xi _i)^{m+1} \right. \nonumber \\
 & & \ \ \ \ \ \ \ \ \ \ \ \ \ \ \ \  \left. +\left( \frac{d}{d\xi _i}{\cal P}^{-1}
\xi _E\right) ^{-1}\left( {\cal P}^{-1} \xi _E\right) ^2 \left( \frac{dV_j^{-1} V_F\Gamma
{\cal P}^{-1}\xi _E}{d\Gamma {\cal P}^{-1}\xi _E}\right) ^2e_n^{pj}\left( V_j^{-1} V_F
\Gamma {\cal P}^{-1}\xi _E\right) ^{n+1}\right] \nonumber \\
 & & +V_1{\cal P}V_2^\dagger \sum_{\scriptstyle p=1\atop
\scriptstyle p\neq \scriptstyle b,c}^{N_1}(-1)^p \prod _{\scriptstyle k=1 \atop
{\scriptstyle k\neq b,c \atop \scriptstyle k\neq p}}^{N_1} \sum_{q,t=-1}^\infty
\left[ \sum_{\scriptstyle l=1 \atop \scriptstyle l\neq E}^{N_1}e_q^{kl} E_{qt}(V_{0l}^{-1}V_l)
b_t^l +\sum_{\scriptstyle l=1 \atop \scriptstyle l\neq F}^{N_2} e_q^{kE}E_{qt}
\left( V_{0E}^{-1}V_E{\cal P} \Gamma V_F^{-1}V_l\right) b_t^l\right] \nonumber \\
 & & \times \sum_{n,m=-1}^\infty  \left[ e_n^{pi}E_{nm}(V_{0i}^{-1}V_i) (\xi _i)^{m+1}-
\left( \frac{d\xi _j}{d\xi _i}\right) ^{-1} e_n^{pE}E_{nm}(V_{0E}^{-1}V_E)(\xi _j)^{m+1}
\right] \nonumber \\
\label{eq4.73}
 & & \times \prod_{\scriptstyle k=1 \atop \scriptstyle k \neq d,g,h}^{N_2}
\sum_{q,t=-1}^\infty \left[ \sum_{\scriptstyle l=1\atop \scriptstyle l\neq F}^{N_2} e_q^{kl}
E_{qt}(V_{0l}^{-1}V_l) b_t^l +\sum_{\scriptstyle l=1\atop \scriptstyle l\neq E}^{N_1} e_q^{kF}
E_{qt} \left( V_{0F}^{-1}V_F \Gamma {\cal P}^{-1}V_E^{-1} V_l\right) b_t^l\right] =0
\end{eqnarray}
where we have called $C_FU_2^{0\dagger }C_2^{-1}\equiv U_2^\dagger $ and $C_FV_2^{0\dagger }
C_2^{-1}\equiv V_2^\dagger $.

We must now extract the ghosts from vertex $U_1$ in the second term of equation (\ref{eq4.73}):
\begin{eqnarray}
U_1{\cal P} & = & V_1\prod_{\scriptstyle k=1\atop \scriptstyle k\neq b,c}^{N_1}
\sum_{q,t=-1}^\infty \left[ \sum_{\scriptstyle l=1\atop \scriptstyle l\neq E}^{N_1}
e_q^{kl}E_{qt}(V_{0l}^{-1}V_l)b_t^l+e_q^{kE}E_{qt}(V_{0E}^{-1}V_E)b_t^E\right] {\cal P}
\nonumber \\
 & = & V_1{\cal P}\prod_{\scriptstyle k=1\atop \scriptstyle k\neq b,c}^{N_1}
\sum_{q,t=-1}^\infty \left[ \sum_{\scriptstyle l=1\atop \scriptstyle l\neq E}^{N_1}
e_q^{kl}E_{qt}(V_{0l}^{-1}V_l)b_t^l+e_q^{kE}E_{qt}(V_{0E}^{-1}V_E{\cal P})b_t^E\right] \ .
\end{eqnarray}
Identifying $b_t^E$ with $b_{-t}^F$ and passing now these ghosts through the conformal
transformation $C_F$, we obtain
\begin{equation}
U_1{\cal P}C_F=V_1{\cal P}C_F\prod_{\scriptstyle k=1\atop \scriptstyle k\neq b,c}^{N_1}
\sum_{q,t=-1}^\infty \left[ \sum_{\scriptstyle l=1\atop \scriptstyle l\neq E}^{N_1}e_q^{kl}
E_{qt}(V_{0l}^{-1}V_l)b_t^l -e_q^{kE}E_{qt}(V_{0E}^{-1}V_E{\cal P}\Gamma V_F^{-1}V_{0F})b_t^F
\right] \ .
\end{equation}
Using now overlap identities (\ref{eq4.65}) for vertex $V_2^\dagger $ and inserting the
conformal transformation $C_2$, we then obtain
\begin{equation}
\label{eq4.74}
U_1{\cal P}C_FV_2^\dagger =V_1{\cal P}C_FV_2^\dagger \prod_{\scriptstyle k=1\atop
\scriptstyle
k\neq b,c}^{N_1}\sum_{q,t=-1}^\infty \left[ \sum_{\scriptstyle l=1\atop \scriptstyle l
\neq E}^{N_1}e_q^{kl}E_{qt}(V_{0l}^{-1}V_l)b_t^l +\sum_{\scriptstyle l=1\atop \scriptstyle l
\neq F}^{N_2}e_q^{kE}E_{qt}\left( V_{0E}^{-1}V_E{\cal P}\Gamma V_F^{-1}V_l\right) b_t^l\right] .
\end{equation}
Substituting (\ref{eq4.74}) into equation (\ref{eq4.73}) we then obtain the overlap identities
between legs $i$ and $j$:
\begin{eqnarray}
 & & U_c\left[ b^i(\xi _i)-\left( \frac{d}{d\xi _i}{\cal P}^{-1}V^{-1}_EV_i\xi _i\right) ^2
\left( {\cal P}^{-1}V_E^{-1}V_i\xi _i\right) ^{-4} \right. \nonumber \\
 & & \ \ \ \ \ \ \ \ \ \ \ \ \ \ \ \ \ \ \ \ \ \ \ \ \ \ \ \ \ \ \ \ \ \ \ \ \ \ \ \ \ \
\left. \times \left( \frac{d}{d\xi _i}{\cal P}^{-1}V_E^{-1}V_i\xi _i\right) ^2 b^j\left(
V_j^{-1}V_F\Gamma {\cal P}^{-1}V_E^{-1}V_i\xi _i\right) \right] =0\ ,\\
 & & U_c\left[ c^i(\xi _i) +\left( \frac{d}{d\xi _i} {\cal P}^{-1}V^{-1}_EV_i\xi _i
\right) ^{-1} \left( {\cal P}^{-1}V_E^{-1} V_i\xi _i\right) ^2 \right. \nonumber \\
 & & \ \ \ \ \ \ \ \ \ \ \ \ \ \ \ \ \ \ \ \ \ \ \ \ \ \ \ \ \ \ \ \ \ \ \ \ \ \ \ \ \ \
\left. \times \left( \frac{d}{d\xi _i} {\cal P}^{-1} V_E^{-1}V_i\xi _i\right) ^{-1} c^j
\left( V_j^{-1} V_F\Gamma {\cal P}^{-1} V_E^{-1}V_i\xi _i\right) \right] \nonumber \\
 & & +V_c \prod _{\scriptstyle k=1 \atop {\scriptstyle k\neq b,c \atop \scriptstyle k
\neq p}}^{N_1} \sum_{q,t=-1}^\infty \left[ \sum_{\scriptstyle l=1 \atop \scriptstyle l
\neq E}^{N_1} e_q^{kl}E_{qt} (V_{0l}^{-1}V_l)b_t^l +\sum_{\scriptstyle l=1\atop
\scriptstyle l\neq F}^{N_2} e_q^{kE}E_{qt} \left( V_{0E}^{-1} V_E{\cal P} \Gamma V_F^{-1}
V_l\right) b_t^l\right] \nonumber \\
 & & \times \sum_{\scriptstyle p=1 \atop \scriptstyle p\neq d,g,h}^{N_2}{-1}^p
\prod_{\scriptstyle k=1 \atop {\scriptstyle k \neq d,g,h \atop \scriptstyle k \neq p}}^{N_2}
\sum_{q,t=-1}^\infty \left[ \sum_{\scriptstyle l=1 \atop \scriptstyle l\neq E}^{N_1} e_q^{kF}
E_{qt} \left( V_{0F}^{-1} V_F \Gamma {\cal P}^{-1} V_E^{-1} V_l \right) b_t^l
+\sum_{\scriptstyle l=1 \atop \scriptstyle l\neq F}^{N_2} e_q^{kl} E_{qt} (V_{0l}^{-1}V_l)
b_t^l\right] \nonumber \\
 & & \times \sum_{n,m=-1}^\infty \left[ e_q^{pF} E_{nm}\left( V_{0F}^{-1} V_F{\cal P}^{-1}
V_E^{-1}V_i\right) (\xi _i)^{m+1} +\left( \frac{d}{d\xi _i}{\cal P}^{-1} V_E^{-1}V_i\xi _i
\right) ^{-1}\left( {\cal P}^{-1}V_E^{-1} V_i\xi _i\right) ^2\right. \nonumber \\
 & & \times \left. \left( \frac{dV_j^{-1} V_F\Gamma {\cal P}^{-1}\xi _E}{d \Gamma {\cal P}^{-1}
\xi _E}\right) ^2 e_n^{pj} E_{nm}(V_{0j}^{-1}V_j) \left( V_j^{-1}V_F\Gamma {\cal P}^{-1}V_E^{-1} V_i\xi _i\right) ^{m+1}\right] \nonumber \\
 & & +V_c\sum_{\scriptstyle p=1\atop \scriptstyle p\neq b,c}^{N_1}(-1)^p \prod _{\scriptstyle
k=1\atop {\scriptstyle k\neq b,c\atop \scriptstyle k\neq p}}^{N_1} \sum_{q,t=-1}^\infty
\left[ \sum_{\scriptstyle l=1\atop \scriptstyle l\neq E}^{N_1} e_q^{kl} E_{qt}(V_{0l}^{-1}V_l)
b_t^l +\sum_{\scriptstyle l=1\atop \scriptstyle l\neq F}^{N_2} e_q^{kE} E_{qt}\left(
V_{0E}^{-1}V_E{\cal P} \Gamma V_F^{-1}V_l\right) b_t^l\right] \nonumber \\
 & & \times \sum_{n,m=-1}^\infty \left[ e_n^{pi} E_{nm}(V_{0i}^{-1}V_i )(\xi _i)^{m+1}
-\left( \frac{d\xi _j}{d\xi _i}\right) ^{-1} e_n^{pE} E_{nm} \left( V_{0E}^{-1}V_j\right)
(\xi _j)^{n+1} \right] \nonumber \\
 & & \times \prod _{\scriptstyle k=1\atop {\scriptstyle k\neq d,g,h\atop \scriptstyle k
\neq p}}^{N_2} \sum_{q,t=-1}^\infty \left[ \sum_{\scriptstyle l=1\atop \scriptstyle l
\neq E}^{N_1} e_q^{kF}E_{qt}\left( V_{0F}^{-1}V_F \Gamma {\cal P}^{-1}V_E^{-1}V_l\right)
b_t^l +\sum_{\scriptstyle l=1\atop \scriptstyle l\neq F}^{N_2} e_q^{kl}E_{qt}(V_{0l}^{-1}
V_l)b_t^l\right] =0\ .
\end{eqnarray}

If we now impose that these are the correct overlap equations between legs $i$ and $j$ of
the composite vertex we then must have:
\begin{equation}
V^{-1}_jV_F\Gamma {\cal P}^{-1}\xi _E=\xi _j
\end{equation}
what fixes the propagator as
\begin{equation}
{\cal P}=V_E^{-1}V_F\Gamma
\end{equation}
which is the same form of the propagator for the bosonic part, but now with the cyclings
defined on the complete generators ${\cal L}_n^i$. The overlap equations now read
\begin{eqnarray}
 & & U_c\left[ b^i(\xi _i)-\left( \frac{d\xi _j}{d\xi _i}\right) ^2b^j(\xi _j)\right] =0\ ,\\
 & & U_c\left[ c^i(\xi _i)-\left( \frac{d\xi _j}{d\xi _i}\right) ^{-1}c^j(\xi _j)\right]
+V_c\prod _{\scriptstyle k=1\atop {\scriptstyle k\neq b,c\atop \scriptstyle k\neq p}}^{N_1}
\sum_{q,t=-1}^\infty \left[ \sum_{\scriptstyle l=1\atop \scriptstyle l\neq E}^{N_1}e_q^{kl}
E_{qt}(V_{0l}^{-1}V_l)b_t^l+\sum_{\scriptstyle l=1\atop \scriptstyle l\neq F}^{N_2}e_q^{kE}
E_{qt}\left( V_{0E}^{-1}V_l\right) b_t^l\right] \nonumber \\
 & & \times \sum_{\scriptstyle p=1\atop \scriptstyle p\neq d,g,h}^{N_2}(-1)^p
\prod _{\scriptstyle k=1\atop {\scriptstyle k\neq d,g,h\atop \scriptstyle k\neq p}}^{N_2}
\sum_{q,t=-1}^\infty \left[ \sum_{\scriptstyle l=1\atop \scriptstyle l\neq E}^{N_1}e_q^{kF}
E_{qt}\left( V_{0F}^{-1}V_l\right) b_t^l+\sum_{\scriptstyle l=1\atop \scriptstyle l\neq F}^{N_2}
e_q^{kl}E_{qt}(V_{0l}^{-1}V_l)b_t^l\right] \nonumber \\
 & & \times \sum_{n,m=-1}^\infty \left[ e_n^{pF}E_{nm}\left( V_{0F}^{-1}V_i\right)
(\xi _i)^{m+1}-\left( \frac{d\xi _j}{d\xi _i}\right) ^{-1}e_n^{pj}E_{nm}(V_{0j}^{-1}V_j)
(\xi _j)^{m+1} \right] \nonumber \\
 & & +V_c\sum_{\scriptstyle p=1\atop \scriptstyle p\neq b,c}^{N_1}(-1)^p\prod _{\scriptstyle
k=1\atop {\scriptstyle k\neq b,c\atop \scriptstyle k\neq p}}^{N_1}\sum_{q,t=-1}^\infty
\left[ \sum_{\scriptstyle l=1\atop \scriptstyle l\neq E}^{N_1}e_q^{kl}E_{qt}(V_{0l}^{-1}V_l)
b_t^l +\sum_{\scriptstyle l=1\atop \scriptstyle l\neq F}^{N_2}e_q^{kE}E_{qt}\left( V_{0E}^{-1}
V_l\right) b_t^l\right] \nonumber \\
 & & \times \sum_{n,m=-1}^\infty \left[ e_n^{pi}E_{nm}(V_{0i}^{-1}V_i)(\xi _i)^{m+1}-\left(
\frac{d\xi _j}{d\xi _i}\right) ^{-1}e_n^{pE}E_{nm}\left( V_{0E}^{-1}V_j\right) (\xi _j)^{n+1}
\right] \nonumber \\
 & & \times \prod _{\scriptstyle k=1\atop {\scriptstyle k\neq d,g,h\atop \scriptstyle k
\neq p}}^{N_2}\sum_{q,t=-1}^\infty \left[ \sum_{\scriptstyle l=1\atop \scriptstyle l
\neq E}^{N_1}e_q^{kF}E_{qt}\left( V_{0F}^{-1}V_l\right) b_t^l+\sum_{\scriptstyle l=1\atop
\scriptstyle l\neq F}^{N_2}e_q^{kl}E_{qt}(V_{0l}^{-1}V_l)b_t^l\right] =0\ .
\end{eqnarray}
These are the overlap identities between legs $i$ and $j$ of the composite vertex $U_c$.

From these overlap equations it is then possible to derive the form of the composite vertex.
It is given by
\begin{eqnarray}
 & & U^{\rm gh}_c=\left( \prod_{\scriptstyle i=1\atop \scriptstyle i\neq E,F}^{N_1+N_2}{ }_i
\langle 0|\right) \exp \left[ \sum_{{\scriptstyle i,j=1\atop \scriptstyle i\neq j}\atop
\scriptstyle i,j\neq E,F}^{N_1+N_2}\sum_{n=2}^\infty \sum_{m=-1}^\infty c_n^iE_{nm}\left(
\Gamma V^{-1}_iV_j\right) b_m^j\right] \times \prod_{r=-1}^1\sum_{\scriptstyle i=1\atop
\scriptstyle i\neq E,F}^{N_1+N_2}\sum_{s=-1}^1E_{rs}(V_i)b_s^i \nonumber \\
 & & \times \prod_{\scriptstyle i=1\atop \scriptstyle i\neq b,c}^{N_1}\sum_{n,m=-1}^\infty
\left[ \sum_{\scriptstyle j=1\atop \scriptstyle j\neq E}^{N_1}e_n^{ij}E_{nm}(V_{0j}^{-1}V_j)
b_m^j
+\sum_{\scriptstyle j=1\atop \scriptstyle j\neq F}^{N_2}e_m^{iE}E_{mn}\left( V_{0E}^{-1}V_j
\right)
b_n^j\right] \nonumber \\
\label{eq4.75}
 & & \times \prod_{\scriptstyle i=1\atop \scriptstyle i\neq d,g,h}^{N_2}\sum_{n,m=-1}^\infty
\left[ \sum_{\scriptstyle j=1\atop \scriptstyle j\neq E}^{N_1}e_n^{iF}E_{nm}\left( V_{0F}^{-1}
V_j\right) b_n^j +\sum_{\scriptstyle j=1\atop \scriptstyle j\neq F}^{N_2}e_n^{ij}E_{nm}
(V_{0j}^{-1}V_j)b_m^j\right] \ .
\end{eqnarray}

Although this is the correct composite vertex for a general cycling, its ghost structure
is not very apparent. We may use the explicit expression for the vectors $e_n^{ij}$ \cite{7}:
\begin{equation}e_n^{ij}=\sum_{m=-1}^\infty k_m^{ij}E_{mn}(\gamma _m^j)\end{equation}
where
\begin{equation}
k_{-1}^{ij}=\delta _{ij}\ \ ,\ \ k_0^{ij}=-\frac{\partial }{\partial z_i}\ln a_0^j\ \ ,\ \
k_n^{ij}=-\frac{\partial }{\partial z_i}\ ,\ n\geq 1
\end{equation}
and
\begin{eqnarray}
\gamma _{-1}^j & = & V_j\ ,\\
\gamma _0^j & = & \exp \left( -{\cal L}_0^l\ln a_0^j\right) \exp \left( -\sum_{n=1}^\infty
\bar a _n^j{\cal L}_n^j\right) \ ,\\
\gamma _p^j & = & \exp \left( -\sum_{n=p+1}^\infty \bar a_n^j{\cal L}_n^j\right) \ \ ,\ \ p
\geq 1\ .
\end{eqnarray}
If we assume now that the cycling transformations $V_E^{-1}$ and $V_F^{-1}$ do not depend on
the variables $z_i$ of vertex $U_1$ or $z_j$ of vertex $U_2^\dagger $, and if we have in mind
that $e_n^{ij}=0$ for any leg $i$ of vertex $U_1$ and a leg $j$ of vertex $U_2^\dagger $ (or
vice-versa), then we have
\begin{equation}e_n^{ij}=\delta ^{ij}\sum_{r=-1}^1E_{-1r}(V_i)\end{equation}
what simplifies things considerably. We shall call all cyclings that have such properties
``simple cycling-like''. In this kind of cyclings, each leg has its own ghost attached to it,
with the exception of three of the legs which have no ghosts attached to them. We will now
consider three cases separately: one in which none of the legs $E$ or $F$ have ghosts attached
to them, one in which one of these legs (say $E$) has a ghost attached to it, and a case where
both legs ($E$ and $F$) have ghosts attached to them.

The composite vertex for the case where neither $E$ nor $F$ (we choose $E=b$ and $F=g$) have
ghosts attached is given by
\begin{equation}
U^{\rm gh}_c=V_c^{\rm gh}\times \prod_{\scriptstyle i=1\atop {\scriptstyle i\neq b\atop
\scriptstyle i\neq E}}^{N_1}\sum_{r=-1}^1E_{-1r}(V_i)b_r^i\times \prod_{\scriptstyle i=1\atop
{\scriptstyle i\neq d,h\atop \scriptstyle i\neq F}}^{N_2}\sum_{r=-1}^1E_{-1r}(V_i)b_r^i
\end{equation}
where $V_c^{\rm gh}$ is the vertex given by the first two terms of equation (\ref{eq4.75}).
For the case where leg $E$ has a ghost attached to it, but not leg $F$ (we shall call $F=g$),
the composite vertex is given by
\begin{equation}
U^{\rm gh}_c = V_c^{\rm gh}\times \prod_{\scriptstyle i=1\atop {\scriptstyle i\neq b,c\atop
\scriptstyle i\neq E}}^{N_1}\sum_{r=-1}^1E_{-1r}(V_i)b_r^i\times \sum_{\scriptstyle j=1\atop
\scriptstyle j\neq F}^{N_2}\sum_{r=-1}^1E_{-1r}(V_j)b_r^j \times \prod_{\scriptstyle i=1\atop
{\scriptstyle i\neq d,h\atop \scriptstyle i\neq F}}^{N_2}\sum_{r=-1}^1E_{-1r}(V_i)b_r^i\ .
\end{equation}
In the last case, where both $E$ and $F$ have ghosts attached to them, we then have
\begin{eqnarray}
U^{\rm gh}_c  & = &  V_c^{\rm gh}\times \prod_{\scriptstyle i=1\atop {\scriptstyle i\neq b,c
\atop \scriptstyle i\neq E}}^{N_1}\sum_{r=-1}^1E_{-1r}(V_i)b_r^i\nonumber \\
 & & \times \sum_{\scriptstyle j=1\atop \scriptstyle j\neq F}^{N_2}\sum_{r=-1}^1E_{-1r}(V_j)
b_r^j \times \sum_{\scriptstyle j=1\atop \scriptstyle j\neq E}^{N_1}\sum_{r=-1}^1E_{-1r}(V_j)
b_r^j\times \prod_{\scriptstyle i=1\atop {\scriptstyle i\neq d,g,h\atop \scriptstyle i
\neq F}}^{N_2}\sum_{r=-1}^1E_{-1r}(V_i)b_r^i\ .
\end{eqnarray}
We can see that, for each case, the composite vertex has the correct ghosts number
($N_1+N_2-2$).

\section{Conclusions}

Using overlap identities, two vertices were sewn together in order to become a composite
vertex. The calculations have been done with the correct ghost numbers for each vertex and the
result has both BRST invariance and the correct ghost counting.

\vskip 0.5 cm

\noindent {\large \bf Acknowledgements}

\vskip 0.3 cm

This work has been done in the Department of Mathematics, King's College London, University of London, under the supervision of Professor P. C. West, and finished at the University of S\~ao Paulo (USP). The author wishes to thank Professor West for suggesting, guiding and correcting this work and Mrs. Nanci Romero for helping with the figures. My gratitude to CAPES and CNPq, Brazil, for financial support. G.E.D.!

\end{document}